\documentclass[11pt,a4paper]{article}
\pdfoutput=1
\usepackage{jcappub}    
\usepackage{amsmath,amssymb}
\usepackage{graphicx}
\usepackage{xcolor}   
\usepackage{tikz}
\usepackage{comment}
\usepackage{subcaption}
\usetikzlibrary{positioning,arrows.meta,backgrounds,calc,fit}

\newcommand{\ftmark}{$^{\star}$}        
\newcommand{\ft}[1]{#1\ftmark}
\newcommand{\SuperSCRIPT}{\texttt{SuperSCRIPT}}
\title{\texttt{SuperSCRIPT}: Reconstructing super-Nyquist ionization structure from coarse reionization lightcones with latent diffusion}

 \author[a,b]{Divesh Jain}                             
 \affiliation[a]{Institut f\"ur Theoretische Physik, Universit\"at Heidelberg,
Philosophenweg 16, 69120 Heidelberg, Germany}                
 \emailAdd{diveshjain.phy@gmail.com}                   

  \author[b,c]{Tirthankar Roy Choudhury}                              
 \affiliation[b]{National Centre for Radio Astrophysics, TIFR,
Pune University Campus, Ganeshkhind, Pune 411007, India}                  
 \affiliation[c]{International Centre for Theoretical Sciences, TIFR,
Bengaluru 560089, India}

  \author[a]{Caroline Heneka}                                                

\abstract{We introduce SuperSCRIPT, a latent-diffusion model for $8\times$
super-resolution of three-dimensional $x_{\rm HII}$ reionization lightcones generated 
using the photon-conserving semi-numerical reionization code \texttt{SCRIPT}. The low-resolution
lightcone specifies the large-scale ionization realization, while four source-model parameters and
the optical depth provide information on the unresolved small-scale morphology for a fixed
matter realization. An anisotropic autoencoder compresses the lightcone into a compact latent space, in which
residual-shift diffusion reconstructs the high-resolution field. At the midpoint of
reionization, where the ionization field is patchiest, the median super-resolved lightcone retains
$97\%$ of the high-resolution cylindrical power at a cross-coherence of $0.90$ within half
the high-resolution Nyquist frequency. The wavelet scattering coefficients of the super-resolved
lightcones agree with the high-resolution ones to within $10\%$ on all but the finest scale.
The autoencoder sets a ceiling to this performance. For $16\times$ super-resolution, finetuning only the diffusion stage reaches the same
ceiling as the $8\times$ model. On a new realization of the matter density field, the frozen
model loses coherence even below the input Nyquist frequency. Finetuning the autoencoder and the diffusion stage for each
realization restores near-ceiling coherence but reduces fidelity on the original realization. Super-resolution takes approximately $4$\,s on one A100 GPU, compared to about $2$
hours for the high-resolution simulation. On a single CPU core, generating the low-resolution
lightcone and super-resolving it is $5.4$ times faster than the high-resolution simulation. Low-resolution ensembles can then be super-resolved as necessary, thus enabling the
generation of high-resolution $21$-cm, kinetic Sunyaev-Zel'dovich (kSZ) and patchy optical
depth maps for parameter estimation and cross-correlation analyses.}
\begin{document}
\maketitle
\flushbottom
\section{Introduction}
\label{sec:intro}

Hydrogen reionization was the last major phase transition of the intergalactic medium (IGM),
determining its subsequent thermal and ionization conditions. Ly$\alpha$ forest observations suggest
a predominantly ionized IGM by $z\simeq5.3$-$6$~\citep{Fan2006, Bosman2022}, with damping-wing
analyses requiring significant neutral hydrogen at $z\gtrsim7$~\citep{Davies2018, Wang2020,
Yang2020} and the CMB optical depth $\tau$ placing the midpoint of reionization near
$z\approx7.7$~\citep{Planck2020}. Observations from JWST further suggest substantial ionizing-photon
budgets that need to be reconciled with these findings~\citep{Munoz2024, Cain2024, Atek2024}. While
these measurements constrain the timing and the volume-averaged ionization history, they leave the
spatial structure of the ionization field underdetermined. This morphology connects galaxy
formation, ionizing-photon production and escape fractions to observables of the IGM. The 21-cm
signal~\citep{Furlanetto2006Review, PritchardLoeb2012}, the
patchy kinetic Sunyaev-Zel'dovich (kSZ) signal~\citep{McQuinn2005, Battaglia2013, Gorce2020,
Jain2024kSZ} and the patchy optical depth~\citep{DvorkinSmith2009, Roy2020, Jain2024tau} all probe
this morphology through the same ionization field. Current 21-cm upper limits already exclude the coldest IGM
models~\citep{HERA2023}, and SKA-Low is designed to map the reionization-era IGM
tomographically~\citep{Koopmans2015}.

Since the density and velocity fields follow from gravity-only simulations, the ionization field is
the one remaining expensive field-level ingredient of these observables, and inference from them
requires repeated forward simulations over volumes several hundred comoving Mpc across that resolve
ionization structure on scales of a few Mpc. Semi-numerical models reduce this cost via approximate
but physically motivated treatments of ionization structure~\citep{Mesinger2011,
ChoudhuryParanjape2018}, yet even these are expensive at the dynamic range required by the observables, when repeated across
parameters and realizations. Most surrogate approaches avoid this cost by emulating compressed
summaries, especially power spectra~\citep{Kern2017, SchmitPritchard2018}, or by applying
simulation-based inference to low-dimensional observables~\citep{Alsing2019, Cranmer2020}. These
methods target prescribed observables and do not provide the fields required for map-level
cross-correlations or additional non-Gaussian statistics. In cosmological large-scale structure
studies, super-resolution has been demonstrated to preserve the large-scale realization of a
low-resolution simulation while reconstructing unresolved matter and halo
structure~\citep{KodiRamanah2020, Li2021, Ni2021, Zhang2023, Zhang2024, Sipp2023}. Diffusion models
for generating cosmological fields~\citep{Ho2020DDPM, MudurFinkbeiner2022} have been used for
conditional super-resolution of large-scale structure~\citep{SchanzListHahn2023, Rouhiainen2024}. In
reionization, deterministic emulators yield realization-specific fields~\citep{Posture2025}, and
conditional diffusion models generate three-dimensional 21-cm brightness-temperature lightcones
across astrophysical parameter space~\citep{Xia2026}. However, the former emulate the 21-cm signal
from a given density field in a coeval box at a single redshift, and the latter generate it
statistically, conditioned on the source parameters alone. Map-level analyses that combine several
observables, such as cross-correlations of the 21-cm, kSZ and optical-depth signals with each other
and with other tracers, instead need the ionization field of a given realization along its full
lightcone, over large volumes at high resolution.

In this work, we introduce \texttt{SuperSCRIPT}, a latent-diffusion model for super-resolution of ionization lightcones generated with the explicitly photon-conserving semi-numerical reionization code \texttt{SCRIPT}~\citep{ChoudhuryParanjape2018}. For lightcones spanning $512\,h^{-1}$cMpc across and $2048\,h^{-1}$cMpc along the line of sight (LOS), over $5\le z\le20$, the model reconstructs ionization fields at a grid resolution of $2\,h^{-1}$cMpc from inputs simulated at $16\,h^{-1}$cMpc. We model the ionization field itself rather than a derived observable, and each reconstruction follows the realization of its low-resolution input, so it can enter cross-correlations and other map-level analyses. The low-resolution field supplies the
resolved structure of the ionization field, and the conditioning parameters specify the
astrophysical source model. From both inputs, the diffusion model reconstructs the unresolved
ionization structure. An anisotropic autoencoder compresses the lightcone according to its
transverse and LOS geometry, enabling residual-shift diffusion~\citep{Yue2023ResShift} to be performed in a smaller latent
volume. We compare each reconstruction with its matched high-resolution simulation using one-point,
spectral and morphological statistics. Super-resolution takes $\sim4$\,s per lightcone on a single
A100 GPU, compared with $\sim2$ hours for the high-resolution simulation. Even on a single CPU core,
generating the low-resolution lightcone and super-resolving it is $5.4$ times faster.

Section~\ref{sec:data} describes the simulations and the dataset, and Section~\ref{sec:method}
presents the super-resolution model. Section~\ref{sec:results} quantifies the reconstruction
fidelity against the matched high-resolution simulations and the computational cost, and tests how
the model generalizes to a higher super-resolution factor and to a new realization of the matter
density field. Finally, Section~\ref{sec:discussion} discusses the results in the context of
map-level, multi-observable analyses of upcoming 21-cm and kSZ surveys and presents our conclusions.

\section{Simulations and Data}
\label{sec:data}

We use the Semi-numerical Code for ReIonization with PhoTon conservation (\texttt{SCRIPT}) to
generate paired high-resolution (HR) and low-resolution (LR) coeval fields of the ionized-hydrogen
fraction, $x_{\rm HII}$, on $256^3$ and $32^3$ grids. The two fields have the same matter
realization and source parameters, while the ionization field is evaluated independently on each
grid. The coeval sequences at each resolution are assembled into paired lightcones.

\subsection{Simulation and source model}
\label{sec:data:sims}

We generate the dark-matter density and velocity fields at $151$ redshifts from $z=20$ to $z=5$ in
steps of $\Delta z=0.1$ with the second-order Lagrangian perturbation theory (2LPT) implementation
in \texttt{MUSIC}~\citep{Hahn2011MUSIC}, using $512^3$ particles at a spacing of $1\,h^{-1}{\rm
cMpc}$ in a periodic box of side $512\,h^{-1}{\rm cMpc}$. Each output is generated directly at its
redshift from the same initial white-noise field. We adopt a flat $\Lambda$CDM cosmology with
$(\Omega_{\rm m},\Omega_\Lambda,\Omega_{\rm b},h,\sigma_8,n_{\rm s},Y_{\rm He})
=(0.308,0.692,0.0482,0.678,0.829,0.961,0.240)$~\citep{Planck2013Parameters}.

The ionization fields are computed with \texttt{SCRIPT}~\citep{ChoudhuryParanjape2018}. Given the matter fields, \texttt{SCRIPT} computes the
halo-collapse field and distributes ionizing photons from source cells, tracking their contribution
to the ionized hydrogen in each cell. This explicit photon counting allows for a numerically
converged large-scale ionization-fluctuation power spectrum across
resolution~\citep{ChoudhuryParanjape2018, Maity2023}.

The source model is characterized by the minimum halo mass hosting ionizing sources, $M_{\rm
min}(z)$, and their ionizing efficiency, $\zeta(z)$. We parameterize their redshift evolution around
the pivot $z_{\rm piv}=8$ by four parameters~\citep{ChoudhuryMukherjeePaul2021, Jain2023framework}:
\begin{align}
\log_{10}\!\left[\frac{M_{\rm min}(z)}{M_\odot}\right]
&= \log_{10}\!\left(\frac{M_{\rm min,0}}{M_\odot}\right)
 + \alpha_M \log_{10}\!\left(\frac{1+z}{1+z_{\rm piv}}\right),\\
\log_{10}\zeta(z)
&= \log_{10}\zeta_0
 + \alpha_\zeta \log_{10}\!\left(\frac{1+z}{1+z_{\rm piv}}\right).
\end{align}
For each choice of parameters, \texttt{SCRIPT} evaluates the bounded ionized-fraction field, $x_{\rm
HII}\in[0,1]$, on the coeval outputs.

\subsection{Lightcone construction}
\label{sec:data:lightcones}

At each redshift, the matter fields are assigned by cloud-in-cell (CIC) to a $256^3$ grid for the HR
evaluation and a $32^3$ grid for the LR evaluation. \texttt{SCRIPT} then evaluates $x_{\rm HII}$
independently on the two grids. The corresponding voxel widths are $\Delta x_{\rm HR}=2\,h^{-1}{\rm
cMpc}$ and $\Delta x_{\rm LR}=16\,h^{-1}{\rm cMpc}$.

The HR and LR lightcones are then assembled from the coeval fields
at each resolution by extracting successive transverse slices at uniform
intervals in comoving distance along the LOS. Each slice is drawn
from the coeval snapshot nearest in redshift, with the slice index wrapped
periodically~\citep{datta2012light}. Each
lightcone is $512\,h^{-1}{\rm cMpc}$ across and $\simeq2048\,h^{-1}{\rm cMpc}$ along the LOS. With
the axes ordered as (LOS, transverse, transverse), the HR reference has shape
$1024\times256\times256$ and the paired LR input has shape $128\times32\times32$. The HR sampling is
therefore eight times finer along each axis. The LOS crosses the periodic simulation volume four
times, so a fixed transverse position returns to the same comoving structure four times along the
LOS, separated by the evolution between their redshifts.

We denote the paired LR input and HR reference by $\mathbf{x}_{\rm
LR}\in\mathbb{R}^{1\times128\times32\times32}$ and $\mathbf{x}_{\rm
HR}\in\mathbb{R}^{1\times1024\times256\times256}$, respectively. The raw-grid Nyquist frequencies
are
\begin{equation}
k_{\rm Nyq}^{\rm LR}=\frac{\pi}{\Delta x_{\rm LR}}
\simeq0.196\,h\,{\rm Mpc}^{-1},
\qquad
k_{\rm Nyq}^{\rm HR}=\frac{\pi}{\Delta x_{\rm HR}}
\simeq1.571\,h\,{\rm Mpc}^{-1}.
\end{equation}

\subsection{Dataset and conditioning vector}
\label{sec:data:dataset}

We sample the source model parameters using a Latin hypercube, keeping
$\log_{10}[M_{\rm min}(z)/M_\odot]\in[7,11]$ and $\log_{10}\zeta(z)\in[0,4]$ at every
$5\le z\le20$, such that 
$\log_{10}(M_{\rm min,0}/M_\odot)\in[7.0,11.0]$, 
$\log_{10}\zeta_0\in[0.0,4.0]$, 
$\alpha_M\in[-7.36,0.0]$, and 
$\alpha_\zeta\in[-7.36,7.36]$. The corresponding LR reionization histories yield Thomson optical
depths of $\tau\in[0.030,0.211]$. The dataset consists of $6000$ training, $1200$ validation and
$600$ test lightcones. The three splits are independent draws over the source-model prior and share
one realization of the initial density field; Section~\ref{sec:results:newseed} tests the model
against a new realization.

The conditional model receives the LR field together with the conditioning vector
\begin{equation}
\mathbf{c}
=
(\log_{10}M_{\rm min,0},\log_{10}\zeta_0,\alpha_M,\alpha_\zeta,\tau),
\end{equation}
whose first four entries are the sampled source-model parameters. The model is described in
Section~\ref{sec:method}.

We define $Q_{\rm HII}(z)$ as the volume-averaged ionized fraction. The Thomson optical depth $\tau$
is obtained by integrating the mass-weighted ionized fraction of the LR lightcone, $\langle x_{\rm
HII}(1+\delta)\rangle$, over redshift. Since \texttt{SCRIPT} conserves photons, the mass of ionized
hydrogen is set by the photon budget, which is the same on both grids, so the values of $\tau$
derived from the LR and HR fields agree to within $0.1\%$.

\section{Method}
\label{sec:method}

\SuperSCRIPT{} learns a conditional distribution of lightcone ionization fields,
$p_\theta(\mathbf{x}_{\rm HR}\mid \mathbf{x}_{\rm LR},\mathbf{c})$, where $\mathbf{c}$ is the
conditioning vector defined in \S\ref{sec:data:dataset}. Given $\mathbf{x}_{\rm LR}$ and $\mathbf{c}$, the model produces the super-resolved (SR) reconstruction $\hat{\mathbf{x}}_{\rm HR}$ of the paired HR field. The LR field supplies the resolved ionization structure, and $\mathbf{c}$ supplies the source model on which the unresolved structure depends.

\begin{figure}
\centering
\includegraphics{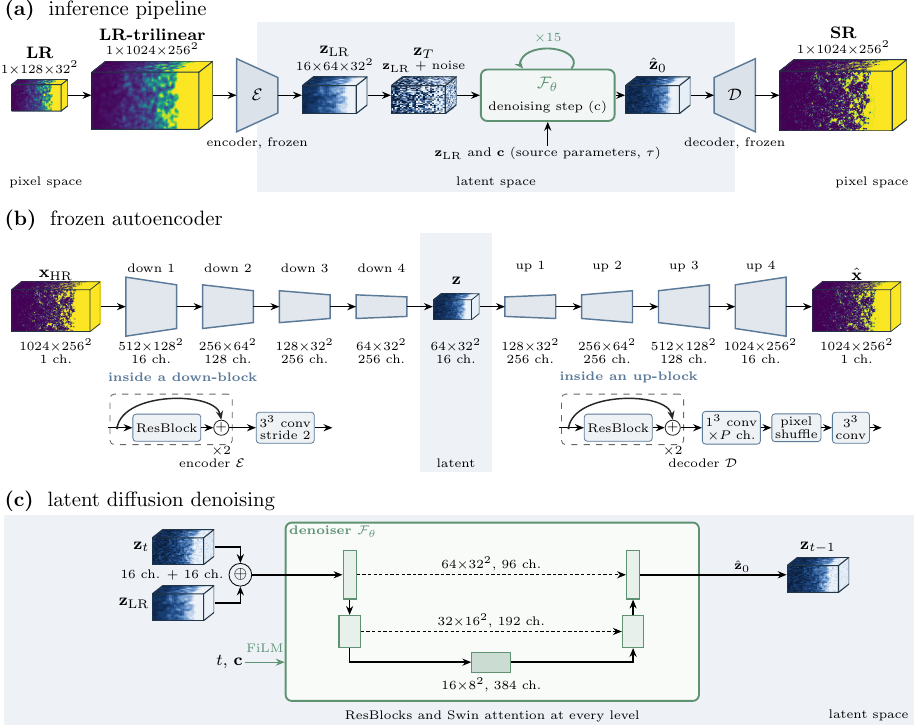}%
{\phantomsubcaption\phantomsection\label{fig:pipeline:a}}%
{\phantomsubcaption\phantomsection\label{fig:pipeline:b}}%
{\phantomsubcaption\phantomsection\label{fig:pipeline:c}}%
\caption{Architecture of the \SuperSCRIPT{} model. \emph{(a)} The LR lightcone is trilinearly
upsampled to the HR grid and encoded and standardized into the LR-conditioning latent
$\mathbf{z}_{\rm LR}$. The chain starts from $\mathbf{z}_T\sim\mathcal{N}(\mathbf{z}_{\rm
LR},\kappa^2\eta_T\mathbf{I})$ and applies $15$ denoising steps, and the decoder maps
$\hat{\mathbf{z}}_0$ to the SR lightcone. \emph{(b)} The autoencoder (Table~\ref{tab:ae-arch}) is
trained on HR fields before the denoiser and then frozen (\S\ref{sec:method:training}). Every block
applies two ResBlocks and then resamples the grid. An encoder block halves it with a stride-2 $3^3$
convolution. A decoder block doubles it by expanding the channels $P$-fold with a $1^3$ convolution,
distributing them over the $P$ finer voxels of each coarse voxel with a pixel shuffle and refining
the result with a $3^3$ convolution. \emph{(c)} In each denoising step, the noisy latent
$\mathbf{z}_t$ is concatenated with $\mathbf{z}_{\rm LR}$ and passed through the denoiser
$\mathcal{F}_\theta$, a three-level U-Net with ResBlocks, Swin attention and skip connections
(Table~\ref{tab:diff-arch}). FiLM feeds the diffusion time $t$ and the source parameters and optical
depth $\mathbf{c}$ into every ResBlock. The denoiser returns an estimate $\hat{\mathbf{z}}_0$ of the
clean latent, from which eq.~\eqref{eq:posterior-mean} gives $\mathbf{z}_{t-1}$, with noise added at
every step except the last.}
\label{fig:pipeline}
\label{fig:ae-architecture}
\end{figure}

Direct diffusion on the $1024\times256\times256$ HR grid would be expensive, as every step of
the reverse chain evaluates the denoiser on every voxel of the grid. We therefore perform the
residual-shift process in a compressed latent space. Residual shifting starts the
reverse chain from the LR latent plus noise, this allows LR to denoise into HR in only $15$ steps~\citep{Yue2023ResShift}. The method consists of two networks trained in
sequence. An anisotropic autoencoder, trained on HR fields and then frozen, defines the latent
(\S\ref{sec:method:ae}), and a denoiser runs the reverse chain in this latent
(\S\ref{sec:method:diffusion}). The full pipeline is outlined in Figure~\ref{fig:pipeline}.

\subsection{Anisotropic autoencoder latent representation}
\label{sec:method:ae}
\label{sec:method:ae:structure}

We represent each HR lightcone with a three-dimensional convolutional autoencoder adapted from
CogVideoX~\citep{Yang2024CogVideoX}, shown in Figure~\ref{fig:pipeline:b}. Its configuration is
specified in Table~\ref{tab:ae-arch}. CogVideoX treats a video as a volume with two spatial and one
temporal coordinate. A reionization lightcone has an analogous geometry, with two periodic
transverse coordinates and a non-periodic LOS coordinate along which redshift evolves. Within each
transverse slice, the ionization field varies strongly between ionized bubbles and neutral regions,
whereas consecutive slices along the LOS have nearly the same statistics, which evolve only slowly
with redshift. The LOS axis therefore tolerates a higher compression than the transverse axes. The diffusion model operates on the resulting compressed representation~\citep{Rombach2022LDM, Esser2021VQGAN}.

With $C=16$ latent channels, the encoder and decoder are the maps
$\mathcal{E}:\mathbb{R}^{1\times1024\times256\times256}
\longrightarrow\mathbb{R}^{16\times64\times32\times32}$ and
$\mathcal{D}:\mathbb{R}^{16\times64\times32\times32}
\longrightarrow\mathbb{R}^{1\times1024\times256\times256}$, respectively. The spatial grid is
contracted by a factor of $16$ along the LOS and $8$ along each transverse axis. The $67\times10^6$
voxels of an HR lightcone are mapped onto $64\times32\times32\simeq65\times10^3$ latent voxels. With $16$ channels in each, the latent holds about $10^6$ values, $64$ times fewer than the HR lightcone.

The HR and LR fields of each pair pass through the same frozen encoder. The HR field gives the
target latent $\mathbf{z}_{\rm HR}=\mathcal{E}(\mathbf{x}_{\rm HR})$, and the LR field, trilinearly
interpolated to the HR grid, gives $\mathcal{E}\!\left(I_{\rm tri}(\mathbf{x}_{\rm LR})\right)$. The
target and the LR latent thus share one $16\times(64\times32\times32)$ latent coordinate system. Since the encoder is trained only on HR fields, $\mathbf{z}_{\rm LR}$ serves as the starting point and conditioning signal from which the denoiser recovers $\mathbf{z}_0$.

Before diffusion training, we standardize each latent channel with its mean and standard deviation
over the training HR latents, so that the noise added in \S\ref{sec:method:diffusion} has the same
relative amplitude in every channel. The same transform is carried out on the HR
and LR latents,
\begin{equation}
\mathbf{z}_0=\operatorname{std}(\mathbf{z}_{\rm HR}),
\qquad
\mathbf{z}_{\rm LR}=\operatorname{std}\!\left[\mathcal{E}\!\left(I_{\rm tri}(\mathbf{x}_{\rm LR})\right)\right] ,
\end{equation}
and its inverse is applied before decoding.

In the encoder, the HR lightcone passes through four compression stages. A stem
$3{\times}3{\times}3$ convolution first maps the ionized fraction to $16$ channels. In each stage,
two residual blocks refine them, each adding the output of two convolutions to its unchanged input,
with group normalization in $16$ groups and a SiLU activation before each
convolution~\citep{Wu2018GroupNorm, Elfwing2018SiLU, He2016PreAct}. A strided convolution then
merges neighbouring voxels, halving every axis in the first three stages and only the LOS in the
last, with a stride of $(2,1,1)$, while the channel count rises to $128$ and then $256$. All
autoencoder convolutions wrap around the periodic transverse axes and replicate the edge slices
along the LOS.
Two further residual blocks act on the latent grid, and the head returns a mean and a log-variance
for each of the $C$ latent channels, as in CogVideoX~\citep{Yang2024CogVideoX}. For training and inference we use the mean, so the encoding is deterministic.

The decoder retraces this path in reverse. A stem convolution expands the latent to $256$ channels
and two residual blocks refine it, after which each of the four up-blocks applies two more residual
blocks and doubles the grid, three of them along all axes and one along the LOS only. Each doubling
is a learned sub-pixel convolution operation~\citep{Shi2016SubPixel}. After the fourth up-block, a convolution maps the $16$ channels back to the ionized fraction. 

\begin{table}
\centering
\small
\setlength{\tabcolsep}{4pt}
\caption{Autoencoder layer specification. The output resolution is ordered (LOS, transverse,
transverse), and the channel column denotes the input$\,\to\,$output width.}
\label{tab:ae-arch}
\begin{tabular}{lccl}
\hline
Stage & Output resolution & Channels & Operation \\
\hline
\multicolumn{4}{l}{\emph{Encoder} $\mathcal{E}$} \\
Input                 & $1024\times256\times256$ & $1$        & --- \\
Stem convolution      & $1024\times256\times256$ & $1\to16$   & $3^3$ conv \\
Down-block 1          & $512\times128\times128$  & $16\to16$  & $2\times$ ResBlock $+$ stride-$2$ conv \\
Down-block 2          & $256\times64\times64$    & $16\to128$ & $2\times$ ResBlock $+$ stride-$2$ conv \\
Down-block 3          & $128\times32\times32$    & $128\to256$& $2\times$ ResBlock $+$ stride-$2$ conv \\
Down-block 4 (LOS)    & $64\times32\times32$     & $256\to256$& $2\times$ ResBlock $+$ stride-$2$ conv \\
Mid-block             & $64\times32\times32$     & $256$      & $2\times$ ResBlock \\
Encoder head          & $64\times32\times32$     & $256\to16{+}16$ & GroupNorm--SiLU $+$ $3^3$ conv \\
\hline
Latent $\mathbf{z}$ (mean) & $64\times32\times32$ & $C=16$     & --- \\
\hline
\multicolumn{4}{l}{\emph{Decoder} $\mathcal{D}$} \\
Stem convolution      & $64\times32\times32$     & $16\to256$ & $3^3$ conv \\
Mid-block             & $64\times32\times32$     & $256$      & $2\times$ ResBlock \\
Up-block 1 (LOS)      & $128\times32\times32$    & $256\to256$& $2\times$ ResBlock $+$ sub-pixel conv \\
Up-block 2            & $256\times64\times64$    & $256\to256$& $2\times$ ResBlock $+$ sub-pixel conv \\
Up-block 3            & $512\times128\times128$  & $256\to128$& $2\times$ ResBlock $+$ sub-pixel conv \\
Up-block 4            & $1024\times256\times256$ & $128\to16$ & $2\times$ ResBlock $+$ sub-pixel conv \\
Output head           & $1024\times256\times256$ & $16\to1$   & GroupNorm--SiLU $+$ $3^3$ conv \\
\hline
\end{tabular}
\end{table}

\subsection{Residual-shift diffusion}
\label{sec:method:diffusion}

We adapt residual-shift diffusion, henceforth ResShift~\citep{Yue2023ResShift}, to the standardized
latents of \S\ref{sec:method:ae} (Figure~\ref{fig:pipeline:c}). It maps the LR latents to the HR
latents, so given $(\mathbf{z}_{\rm LR},\mathbf{c})$ the reverse chain only has to supply the
residual $\mathbf{z}_0-\mathbf{z}_{\rm LR}$.

For diffusion step $t\in\{1,\ldots,T\}$, the forward process samples
\begin{equation}
\mathbf{z}_t
=
(1-\eta_t)\mathbf{z}_0
+
\eta_t\mathbf{z}_{\rm LR}
+
\kappa\sqrt{\eta_t}\,\boldsymbol{\varepsilon},
\qquad
\boldsymbol{\varepsilon}\sim\mathcal{N}(\mathbf{0},\mathbf{I}) .
\label{eq:resshift-forward}
\end{equation}
Here $\{\eta_t\}_{t=1}^{T}$ is a monotone shifting schedule and $\kappa$ determines the noise
amplitude. As $t$ increases, the mean of $\mathbf{z}_t$ shifts from the clean latent $\mathbf{z}_0$
at $\eta_1\simeq0$ to the LR latent $\mathbf{z}_{\rm LR}$ at $\eta_T\simeq1$, while the noise
variance grows to $\kappa^2\eta_T$. The reverse chain therefore starts from
\[
\mathbf{z}_T
\simeq
\mathbf{z}_{\rm LR}+\kappa\sqrt{\eta_T}\,\boldsymbol{\varepsilon} .
\]
We set $\kappa=2$ and use a chain of $T=15$ steps with an exponential schedule.

The corresponding reverse posterior has a Gaussian mean
\begin{equation}
\boldsymbol{\mu}_t(\mathbf{z}_t,\mathbf{z}_0)
=
\frac{\eta_{t-1}}{\eta_t}\mathbf{z}_t
+
\frac{\eta_t-\eta_{t-1}}{\eta_t}\mathbf{z}_0 ,
\label{eq:posterior-mean}
\end{equation}
and a variance fixed by the ResShift schedule~\citep{Yue2023ResShift}. At inference, the denoiser
predicts
\[
\hat{\mathbf{z}}_0
=
\mathcal{F}_{\theta}(\mathbf{z}_t,t,\mathbf{z}_{\rm LR},\mathbf{c})
\]
and is used in Eq.~\eqref{eq:posterior-mean}. We sample
$\mathbf{z}_{t-1}$ from the resulting Gaussian, adding fresh noise at every denoising step except the final
step.

For each training pair, we sample a diffusion step $t$, draw $\boldsymbol{\varepsilon}\sim\mathcal{N}(\mathbf{0},\mathbf{I})$,
construct $\mathbf{z}_t$ from Eq.~\eqref{eq:resshift-forward}, and minimize
\begin{equation}
\mathcal{L}_{\rm diff}(\theta)
=
\mathbb{E}_{\mathbf{z}_0,\mathbf{z}_{\rm LR},t,\boldsymbol{\varepsilon}}
\left[
\left\|
\mathbf{z}_0
-
\mathcal{F}_{\theta}(\mathbf{z}_t,t,\mathbf{z}_{\rm LR},\mathbf{c})
\right\|_2^2
\right].
\label{eq:diff-loss}
\end{equation}

We discuss the conditioning mechanism for the denoiser in (\S\ref{sec:method:cond}) and its network architecture in (\S\ref{sec:method:diffusion:unet}).

\subsubsection{Conditioning the denoiser}
\label{sec:method:cond}

The denoiser takes in the spatial latent $\mathbf{z}_{\rm LR}$ and the global conditioning vector
$\mathbf{c}$ via separate pathways.

\paragraph{Spatial conditioning.}
\label{sec:method:cond:lrcat}

We concatenate the noisy latent $\mathbf{z}_t$ with the LR-conditioning latent $\mathbf{z}_{\rm LR}$
along the channel axis, which keeps the LR field in view of the denoiser at every step. As they lie
on the same $64\times32\times32$ latent grid, they stay voxel-aligned.

\paragraph{Global parameter conditioning.}
\label{sec:method:cond:filmcond}

The pivot parameters $\log_{10}M_{\rm min,0}$ and $\log_{10}\zeta_0$ specify the source model at
at $z_{\rm piv}=8$, and the slopes $(\alpha_M,\alpha_\zeta)$ govern its redshift evolution.
The optical depth $\tau$ provides the integrated reionization history. Every lightcone is generated
with a unique combination of source parameters, which $\mathbf{c}$ supplies to every stage of the
denoiser. Each component of this conditioning vector $\mathbf{c}$ is mapped
linearly from its training range in \S\ref{sec:data:dataset} to $[-1,1]$, and the rescaled vector is passed through a two-layer multilayer perceptron (MLP) of width $384$,
four times the base width of the denoiser, and added to the sinusoidal diffusion-time embedding.
\begin{equation}
\mathbf{e}(t,\mathbf{c})
=
\mathrm{MLP}_t(\mathrm{sin\text{-}embed}(t))
+
\mathrm{MLP}_{\mathbf{c}}(\mathbf{c}) .
\label{eq:emb-sum}
\end{equation}
Each residual block then projects this combined embedding to a feature-wise linear modulation
(FiLM) scale and shift of its group-normalized activations~\citep{Perez2018FiLM,
Dhariwal2021DiffusionBeatsGANs}.

\subsubsection{Denoiser architecture}
\label{sec:method:diffusion:unet}

The denoiser $\mathcal{F}_\theta$ is a three-dimensional
U-Net~\citep{Ronneberger2015UNet,Cicek20163DUNet} with Swin-transformer
self-attention over local windows~\citep{Liu2021Swin}, adapted from the \texttt{UNetModelSwin} of Yue
et al.~\citep{Yue2023ResShift}. Its configuration is specified in Table~\ref{tab:diff-arch}. A
$3^3$ convolution expands the $32$-channel input to the base width of $96$ channels, after
$\mathbf{z}_t$ is divided by $\sqrt{1+\kappa^2\eta_t}$ to keep its variance near unity at every
step. The encoder has three stages with channel multipliers $(1,2,4)$, which widen
$96\to192\to384$ and halve each spatial axis $(64,32,32)\to(32,16,16)\to(16,8,8)$. Each stage
applies three residual blocks and Swin self-attention in windows of $4^3$ latent voxels. Here, each
residual block applies GroupNorm--SiLU--$3^3$ conv twice and adds the result to its input, and the
FiLM scale and shift of \S\ref{sec:method:cond:filmcond} act on the second GroupNorm. The
bottleneck then applies Swin self-attention between two residual blocks. The decoder mirrors the
encoder, with a skip connection
from each encoder stage, and produces the $16$-channel prediction $\hat{\mathbf{z}}_0$. The full
backbone has $1.2\times10^{8}$ parameters.

\begin{table}
\centering
\small
\setlength{\tabcolsep}{4pt}
\caption{Denoising backbone $\mathcal{F}_\theta$ layer specification. The output resolution is
ordered (LOS, transverse, transverse), and the channel column denotes the input$\,\to\,$output width.}
\label{tab:diff-arch}
\begin{tabular}{lccl}
\hline
Stage & Output resolution & Channels & Operation \\
\hline
Input concatenation   & $64\times32\times32$ & $16{+}16\to32$ & concatenate $\mathbf{z}_t$, $\mathbf{z}_{\rm LR}$ \\
Stem convolution      & $64\times32\times32$ & $32\to96$      & $3^3$ conv \\
\multicolumn{4}{l}{\emph{Encoder}} \\
Stage 1               & $32\times16\times16$ & $96\to96$      & $3\times$ ResBlock $+$ Swin attn. $+$ stride-$2$ conv \\
Stage 2               & $16\times8\times8$   & $96\to192$     & $3\times$ ResBlock $+$ Swin attn. $+$ stride-$2$ conv \\
Stage 3               & $16\times8\times8$   & $192\to384$    & $3\times$ ResBlock $+$ Swin attn. \\
Bottleneck            & $16\times8\times8$   & $384$          & ResBlock $+$ Swin attn. $+$ ResBlock \\
\hline
\multicolumn{4}{l}{\emph{Decoder} (with U-Net skips)} \\
Stage 3$'$            & $32\times16\times16$ & $384\to384$    & $4\times$ ResBlock $+$ Swin attn. $+$ upsample \\
Stage 2$'$            & $64\times32\times32$ & $384\to192$    & $4\times$ ResBlock $+$ Swin attn. $+$ upsample \\
Stage 1$'$            & $64\times32\times32$ & $192\to96$     & $4\times$ ResBlock $+$ Swin attn. \\
Output head           & $64\times32\times32$ & $96\to16$      & GroupNorm--SiLU $+$ $3^3$ conv \\
\hline
\end{tabular}
\end{table}

\subsection{Two-stage training}
\label{sec:method:training}

We first train the autoencoder on HR fields and freeze its exponential moving average (EMA) weights.
Then, we train the denoiser on cached standardized latents. The frozen representation fixes the
latent coordinate system and sets the ceiling for the reconstruction fidelity for the diffusion
model.

\paragraph{Autoencoder training.}

The autoencoder is trained to reconstruct bounded HR ionization fields. For an HR field
$\mathbf{x}_{\rm HR}$ and a reconstruction $\hat{\mathbf{x}}=\mathcal{D}(\mathcal{E}(\mathbf{x}_{\rm
HR}))$, the reconstruction loss is given by
\begin{equation}
\mathcal{L}_{\rm recon}
=
\tfrac{1}{2}\left\|\hat{\mathbf{x}}-\mathbf{x}_{\rm HR}\right\|_2^2
+
\tfrac{1}{2}\left\|\hat{\mathbf{x}}-\mathbf{x}_{\rm HR}\right\|_1
+
\lambda_{\rm b}\left\|\operatorname{dist}\big(\hat{\mathbf{x}},[0,1]\big)\right\|_2^2 
\label{eq:loss-recon}
\end{equation}
The first two terms are voxelwise $L_2$ and $L_1$ reconstruction losses. The third term, with
$\lambda_{\rm b}=5\times10^{-2}$, penalizes the squared distance of each reconstructed voxel from the
physical range $[0,1]$ of $x_{\rm HII}$. After reconstruction training, the autoencoder is adversarially finetuned against a
three-dimensional PatchGAN discriminator~\citep{Isola2017Pix2Pix}. The discriminator is trained with a
binary cross-entropy loss to separate HR fields from reconstructions, while the autoencoder minimizes
$\mathcal{L}_{\rm recon}+\lambda_{\rm adv}\mathcal{L}_{\rm adv}$ with $\lambda_{\rm adv}=2.5\times10^{-3}$.

\paragraph{Diffusion training.}
With the trained autoencoder, we cache paired latents $(\mathbf{z}_0,\mathbf{z}_{\rm LR})$ for the dataset and train the denoiser with the loss described in Eq.~\eqref{eq:diff-loss}.

\subsection{Inference}
\label{sec:method:inference}
\label{sec:method:inference:pipeline}

At inference, the LR field is encoded and standardized,
$\mathbf{z}_{\rm LR}
=
\operatorname{std}\!\left[\mathcal{E}\!\left(I_{\rm tri}(\mathbf{x}_{\rm LR})\right)\right]$
and the $T=15$-step reverse chain of \S\ref{sec:method:diffusion} runs from $\mathbf{z}_T$,
conditioned on $(\mathbf{z}_{\rm LR},\mathbf{c})$, with the EMA weights of the denoiser. The final
latent is unstandardized and decoded,
$
\hat{\mathbf{x}}_{\rm HR}
=
\mathcal{D}\!\bigl(\operatorname{std}^{-1}(\hat{\mathbf{z}}_0)\bigr)
$.

\section{Results}
\label{sec:results}

We evaluate the model on the held-out cohort of $600$ test lightcones described in
Section~\ref{sec:data}. One-point and higher-order statistics compare SR with HR in real space,
while cylindrical auto-power spectra $P(k_\parallel,k_\perp)$ separate the recovery of LOS and
transverse modes. The SR-to-HR cylindrical power ratio $P_{\rm SR}/P_{\rm HR}$ measures amplitude
recovery, and the SR-HR cross-coherence, $r=P_{\rm SR,HR}/\sqrt{P_{\rm SR}P_{\rm HR}}$, measures
phase recovery. Along both $k_\parallel$ and $k_\perp$ the Nyquist frequency is $k_{\rm Nyq}^{\rm LR}=0.196\,h\,{\rm Mpc}^{-1}$
for the LR grid and $k_{\rm Nyq}^{\rm HR}=1.571\,h\,{\rm Mpc}^{-1}$ for the HR grid
(\S\ref{sec:data:lightcones}). We quote cylindrical statistics within half the HR Nyquist frequency,
$k_\parallel,k_\perp\le k_{\rm Nyq}^{\rm HR}/2=0.785\,h\,{\rm Mpc}^{-1}$. Since the decoder has no
output clamp by design, it can predict ionized fractions $\tilde{x}_{\rm HII}$ outside the physical
range $[0,1]$. For each lightcone we take the rms excursions below $0$ and above $1$, $\sigma_\pm\ge0$, as the
decoder scatter at each limit, rescale the output to $(\tilde{x}_{\rm HII}-\sigma_-)/(1-\sigma_+-\sigma_-)$
and clamp the result to $[0,1]$. We present this
normalized reconstruction unless stated otherwise.

We first discuss the fidelity of reconstructions for selected test lightcones by comparing them
directly with their HR targets (\S\ref{sec:results:qual}). We then quantify population-level
recovery across one-point, morphological and spectral statistics
(\S\ref{sec:results:pdf}--\ref{sec:results:pk}). Next, we examine the dependence of reconstruction
fidelity on the source-model parameters (\S\ref{sec:results:cond}) and report the computational cost
of the reconstruction (\S\ref{sec:results:cost}). Finally, we test model generality by carrying out
$16\times$ super-resolution (\S\ref{sec:results:x16}) and by evaluating the $8\times$ reconstruction
for a different choice of matter-density seed (\S\ref{sec:results:newseed}).

\subsection{Reconstruction fidelity of individual lightcones}
\label{sec:results:qual}

We evaluate reconstructions for four lightcones selected from the test set with optical depths
$\tau=0.040$, $0.054$, $0.081$ and $0.102$, chosen to span different reionization histories. As each
of these histories evolves over different redshift ranges, a common redshift comparison samples
different stages of reionization. We therefore compare the HR targets and SR reconstructions at
fixed ionization stages, $Q_{\rm HII}=0.1$, $0.25$, $0.5$, $0.75$ and $0.9$. We define $z_{q}^{\rm
HR}$ as the redshift at which the volume-averaged ionized fraction $Q_{\rm HII}$ of the HR lightcone
reaches $q\%$. The results below are for the $\tau=0.054$ lightcone, which has the Planck
best-fit optical depth~\citep{Planck2020}. We put them in context with the recovery of the other
three, and Appendix~\ref{app:lightcones} provides their reference figures.

\begin{figure}
\centering
\includegraphics[width=\columnwidth]{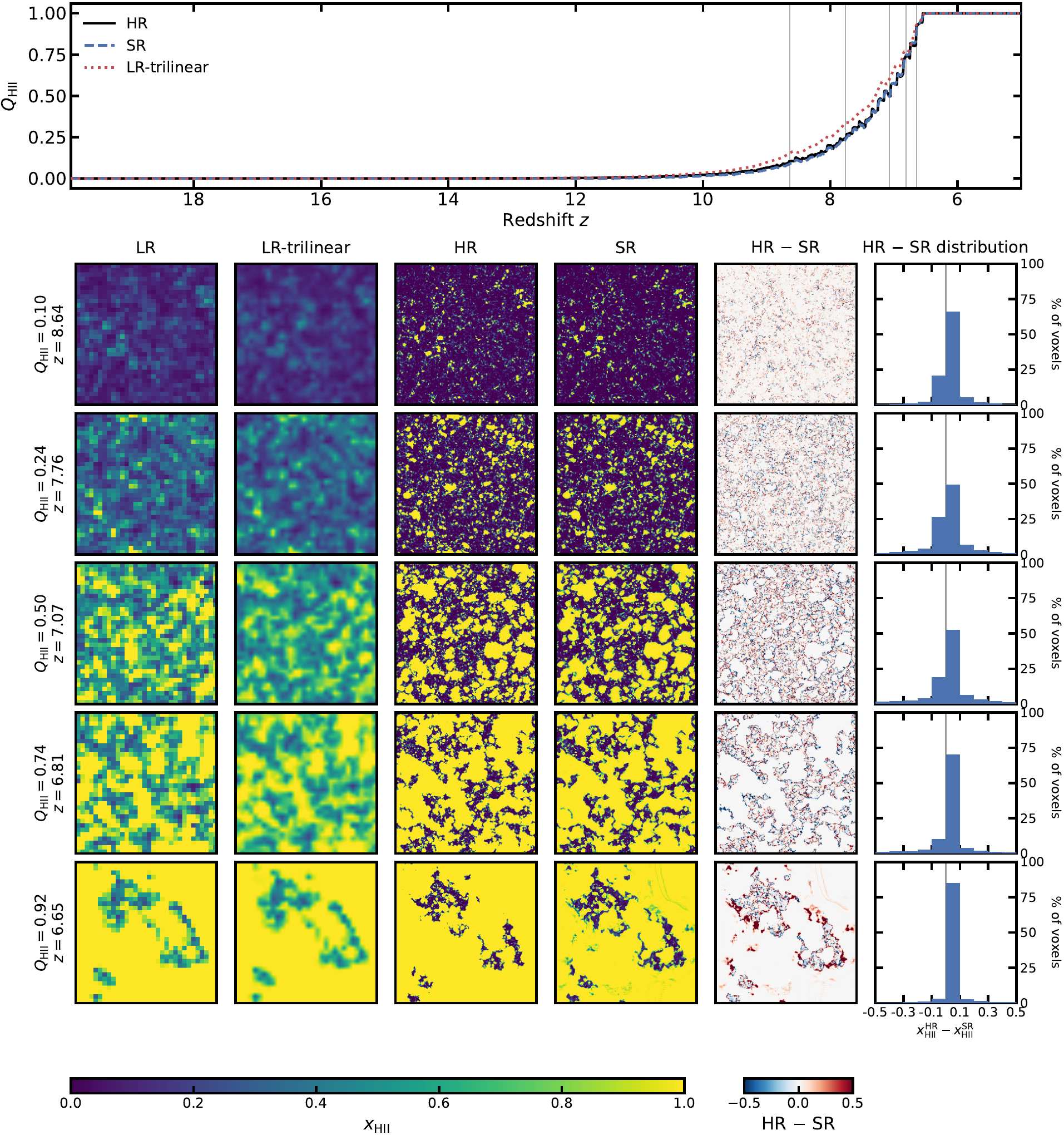}
\caption{Reconstruction of the test lightcone with a Planck-like integrated optical depth
$\tau=0.054$ and midpoint of reionization $z_{50}^{\rm HR}=7.07$. The top panel shows the ionization
histories $Q_{\rm HII}(z)$ of HR, SR and LR-trilinear, with vertical lines marking the redshifts of
the HR slices closest to $Q_{\rm HII}=0.1$, $0.25$, $0.5$, $0.75$ and $0.9$, at $z=8.64$, $7.76$,
$7.07$, $6.81$ and $6.65$. The bottom panel shows the corresponding transverse slices of LR,
LR-trilinear, HR and SR, followed by the residual $x_{\rm HII}^{\rm HR-SR}=x_{\rm HII}^{\rm
HR}-x_{\rm HII}^{\rm SR}$ and its histogram.}
\label{fig:lc:planck}
\end{figure}

Figure~\ref{fig:lc:planck} shows the resulting transverse slices for the $\tau=0.054$ lightcone.
LR retains the coarse, large-scale ionization pattern, and
LR-trilinear maps the LR pattern smoothly onto the HR grid. Both lack the small bubbles and sharp
bubble boundaries that the HR target resolves. SR recovers the HR bubble locations, sizes and
connectivity. The residual $x_{\rm HII}^{\rm HR-SR}\equiv x_{\rm HII}^{\rm HR}-x_{\rm HII}^{\rm SR}$
remains low-amplitude and diffuse through $Q_{\rm HII}\simeq0.5$. At $Q_{\rm HII}>0.5$, the residual
concentrates around bubble boundaries and the remaining neutral islands. The residual histograms
show that the majority of the voxels of each slice differ by at most $0.1$. Near the end of
reionization, SR contains faint small-scale neutral structure in otherwise ionized regions without
systematically biasing $Q_{\rm HII}(z)$. In the reionization history panel the LR-trilinear $Q_{\rm
HII}(z)$, the red dotted curve, rises earlier than HR and SR throughout the transition. This is
because the coarse grid smooths out the overdense regions in which ionization begins, so the same
ionized mass fills a larger volume and the volume-averaged $Q_{\rm HII}$ is higher, while the
mass-weighted ionized fraction that sets $\tau$ is the same on both grids. For reference, we
present the recovery of the mass-weighted ionized fraction in Figure~\ref{fig:qmass}.

\begin{figure}
\centering
\includegraphics[width=\columnwidth]{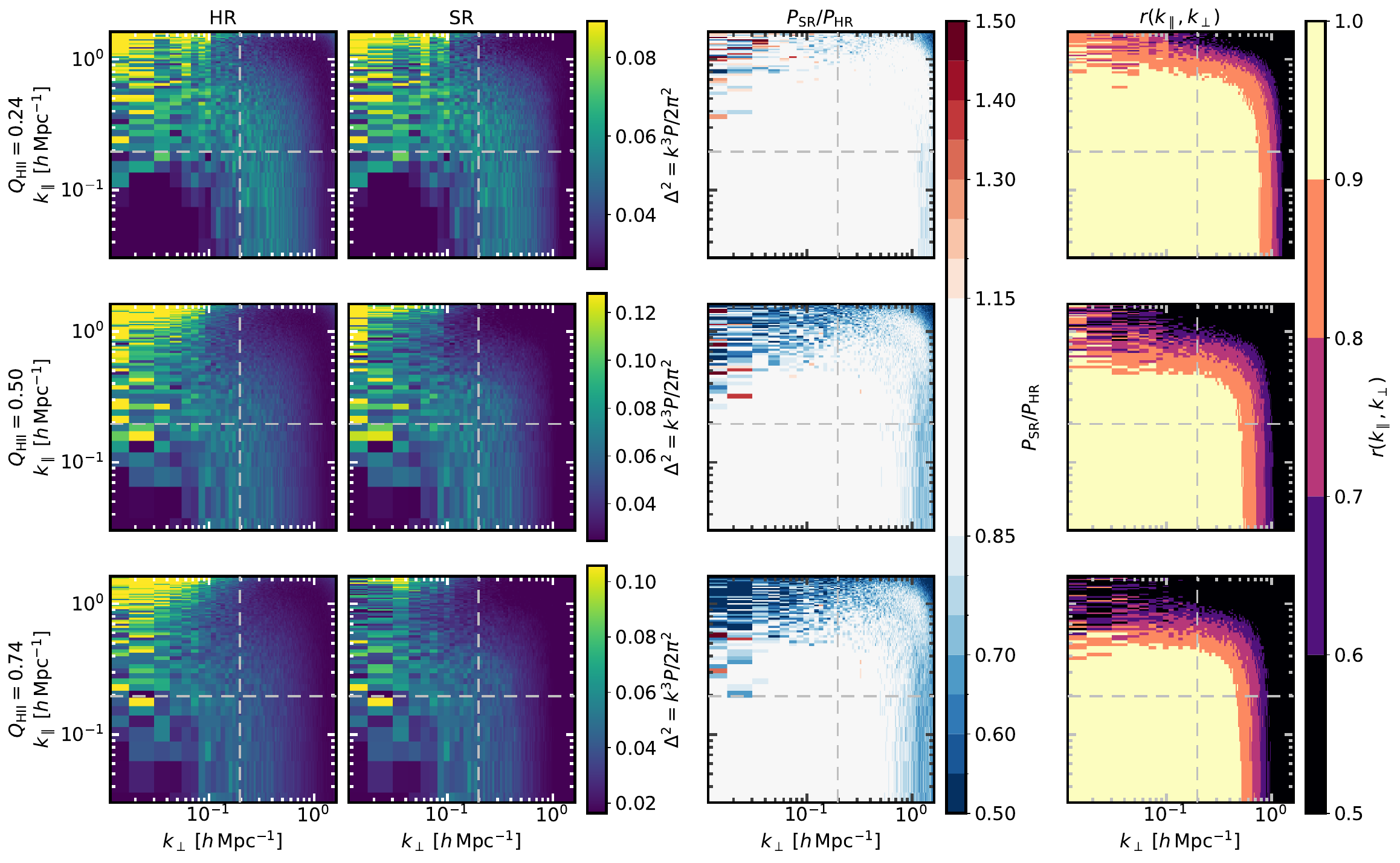}
\caption{Cylindrical spectra of the $\tau=0.054$ lightcone at three stages of reionization.
\emph{Rows:} redshift windows $w_z=1$, labelled by the $Q_{\rm HII}$ of their central HR slice.
\emph{Columns:} the dimensionless cylindrical power $\Delta^2=k^3P(k_\parallel,k_\perp)/(2\pi^2)$,
with $k=(k_\parallel^2+k_\perp^2)^{1/2}$, of the HR and SR fields, their power ratio and their
cross-coherence. White cells mark $0.85\le P_{\rm SR}/P_{\rm HR}\le1.15$. Dashed lines mark the LR
Nyquist frequency $k_{\rm Nyq}^{\rm LR}=\pi/\Delta x_{\rm LR}\simeq0.196\,h\,{\rm Mpc}^{-1}$.}
\label{fig:cyl:s2}
\end{figure}

In Figure~\ref{fig:cyl:s2} we evaluate the dimensionless cylindrical power of HR and SR over redshift windows of width $w_z=1$ centred on $Q_{\rm HII}=0.25$, $0.50$ and $0.75$. These windows retain long-wavelength LOS modes while resolving the
transverse spectrum. SR reconstructs modes beyond the grey dashed LR Nyquist lines along both axes
and is consistent with HR across much of the plane within half the HR Nyquist frequency. The
power-ratio panels quantify the amplitude recovery, with white cells marking $0.85\le P_{\rm
SR}/P_{\rm HR}\le1.15$, while the cross-coherence measures phase recovery. The reconstruction is
consistent with HR to within $15\%$ across most of the plane within half the HR Nyquist
frequency. We therefore summarize the cylindrical power spectrum of each window by one summary
statistic for each quantity, the median over cells of the power ratio and of the coherence.
To put the $\tau=0.054$ lightcone in the context of the other recoveries,
Table~\ref{tab:cyl:showcase} lists these medians at each stage for all four lightcones and for the
test cohort, within half the HR Nyquist frequency and up to $1\,h\,{\rm Mpc}^{-1}$. Within half the HR Nyquist frequency, the median power ratio lies between $0.87$ and
$1.03$ and the coherence between $0.79$ and $0.97$ at every stage in all four lightcones. Extending
the band to $1\,h\,{\rm Mpc}^{-1}$ slightly lowers the power ratio to $0.83$-$0.99$ and the
coherence to $0.68$-$0.94$.

\begin{table}
\centering
\small
\setlength{\tabcolsep}{4pt}
\caption{Recovery of the four selected lightcones at three stages of reionization. $z_{50}^{\rm HR}$
is the redshift at which the HR history reaches $Q_{\rm HII}=0.5$, and $\Delta z^{\rm HR}$ is its
duration, the redshift interval between $Q_{\rm HII}=0.25$ and $0.75$. For each stage, $R_P$ is the
median cylindrical power ratio $P_{\rm SR}/P_{\rm HR}$ and $r$ the median SR--HR cross-coherence
over $k_\parallel,k_\perp\le k_{\max}$, with $k_{\max}=k_{\rm Nyq}^{\rm HR}/2=0.785\,h\,{\rm Mpc}^{-1}$
or $1\,h\,{\rm Mpc}^{-1}$, in the redshift windows of width $w_z=1$ centred on each stage in the
four selected lightcones. The test-cohort rows give the median and $1\sigma$ range of $R_P$ and $r$
over the $452$, $445$ and $401$ test lightcones that reach each stage with the full $w_z=1$ window
inside the lightcone.}
\label{tab:cyl:showcase}
\begin{tabular}{c cc c cc cc cc}
\hline
 & & & & \multicolumn{2}{c}{$Q_{\rm HII}=0.25$} & \multicolumn{2}{c}{$Q_{\rm HII}=0.50$} & \multicolumn{2}{c}{$Q_{\rm HII}=0.75$} \\
$\tau$ & $z_{50}^{\rm HR}$ & $\Delta z^{\rm HR}$ & $k_{\max}$ & $R_P$ & $r$ & $R_P$ & $r$ & $R_P$ & $r$ \\
\hline
$0.040$ & $5.91$ & $0.70$ & $0.785$ & $1.03$ & $0.92$ & $0.96$ & $0.83$ & $0.93$ & $0.79$ \\
 & &  & $1$ & $0.99$ & $0.85$ & $0.91$ & $0.73$ & $0.88$ & $0.68$ \\
$0.054$ & $7.07$ & $0.95$ & $0.785$ & $0.99$ & $0.94$ & $0.93$ & $0.84$ & $0.90$ & $0.81$ \\
 & &  & $1$ & $0.97$ & $0.89$ & $0.89$ & $0.74$ & $0.85$ & $0.68$ \\
$0.081$ & $9.54$ & $1.42$ & $0.785$ & $0.96$ & $0.97$ & $0.95$ & $0.91$ & $0.90$ & $0.85$ \\
 & &  & $1$ & $0.95$ & $0.94$ & $0.90$ & $0.83$ & $0.83$ & $0.73$ \\
$0.102$ & $11.48$ & $1.36$ & $0.785$ & $0.97$ & $0.95$ & $0.93$ & $0.87$ & $0.87$ & $0.81$ \\
 & &  & $1$ & $0.96$ & $0.90$ & $0.90$ & $0.78$ & $0.84$ & $0.69$ \\
\hline
\multicolumn{3}{l}{Test cohort}  & $0.785$ & $1.01^{+0.07}_{-0.05}$ & $0.96^{+0.01}_{-0.03}$ & $0.97^{+0.08}_{-0.05}$ & $0.90^{+0.03}_{-0.04}$ & $0.90^{+0.07}_{-0.05}$ & $0.81^{+0.02}_{-0.04}$ \\
 & &  & $1$ & $1.00^{+0.07}_{-0.06}$ & $0.91^{+0.02}_{-0.04}$ & $0.93^{+0.09}_{-0.05}$ & $0.81^{+0.04}_{-0.06}$ & $0.84^{+0.06}_{-0.04}$ & $0.67^{+0.03}_{-0.04}$ \\
\hline
\end{tabular}
\end{table}

\subsection{Population-level recovery: one-point and morphological statistics}
\label{sec:results:pdf}

We now quantify fidelity over the full test cohort, the $600$ test lightcones of
\S\ref{sec:data:dataset}. The cohort spans reionization histories from early to late, with optical depths ranging from
$\tau=0.03$ to $0.21$. For the $410$ lightcones whose HR history passes through both
$Q_{\rm HII}=0.25$ and $0.75$ inside the lightcone, let $z_{\rm dev}$ be the deviation of the
redshift at which the SR lightcone reaches a given $Q_{\rm HII}$ from that of HR.
For SR the median $z_{\rm dev}$ is $-0.10^{+0.08}_{-0.16}$, $-0.02^{+0.02}_{-0.07}$ and
$0.00^{+0.02}_{-0.02}$ at $Q_{\rm HII}=0.25$, $0.5$ and $0.75$, so SR follows the HR history
closely. Because the optical depth integrates the ionization history, SR also recovers it, with
$\Delta\tau=\tau_{\rm SR}-\tau_{\rm HR}=-0.0013^{+0.0006}_{-0.0013}$ over the $600$ test lightcones, well
within the $1\sigma$ uncertainty of $0.007$ of the Planck measurement~\citep{Planck2020}.

\begin{figure}
\centering
\includegraphics[width=\columnwidth]{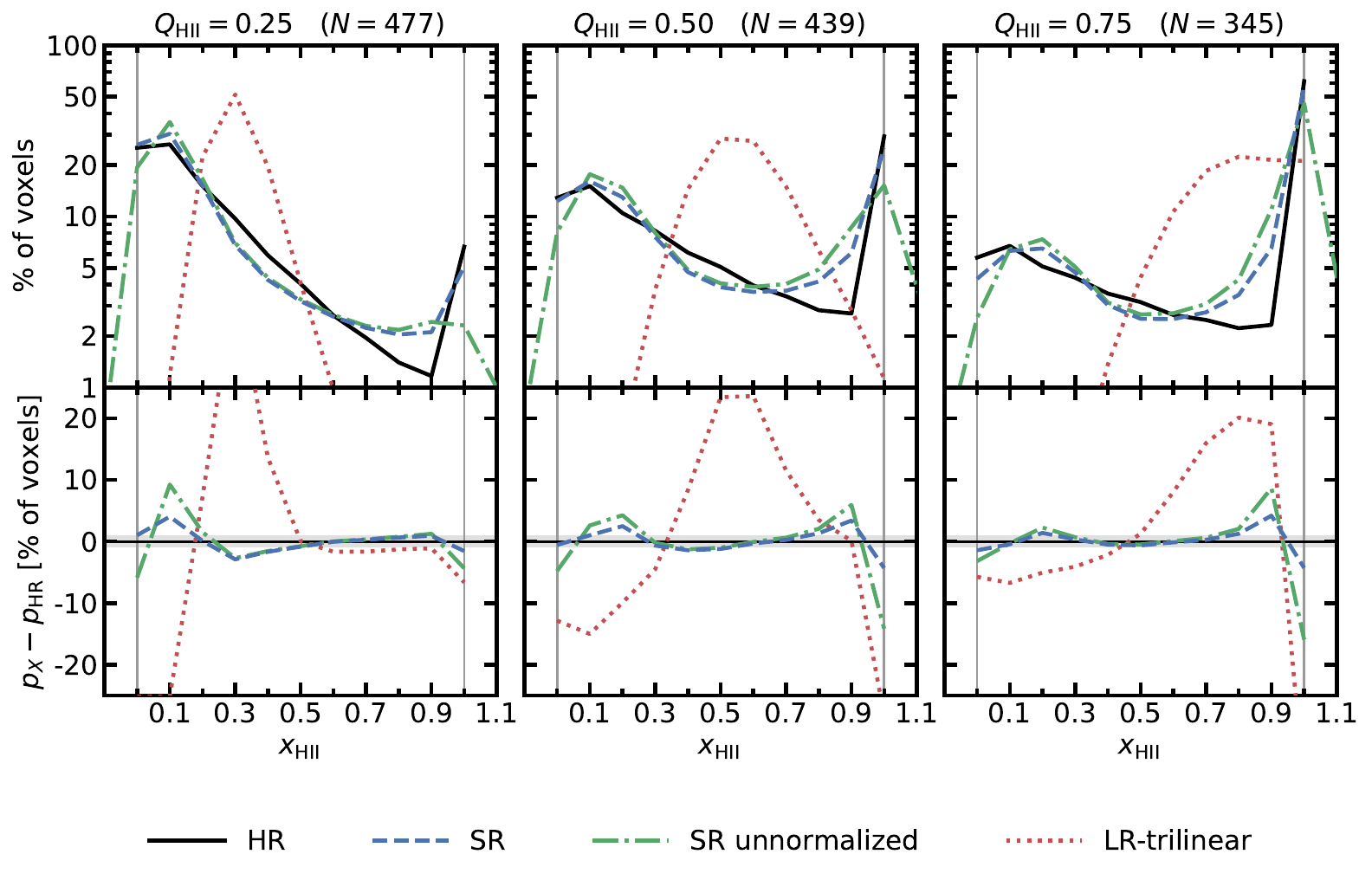}
\caption{Voxel distribution of $x_{\rm HII}$ at three matched ionization stages, as the percentage
of voxels per bin of width $0.1$. The lower panels show the difference of each distribution from
that of HR, $p_X-p_{\rm HR}$, in the same unit. SR unnormalized is the decoder output before the
normalization; $N$ is the number of lightcones that reach each stage. Vertical lines mark the
physical bounds $x_{\rm HII}=0$ and $1$.}
\label{fig:voxel}
\end{figure}

\paragraph{Voxel distribution.}
We compare voxel distributions of the test cohort at three matched ionization stages, $Q_{\rm
HII}=0.25$, $0.5$ and $0.75$. In each lightcone we select the slices at which the HR history reaches
the stage and evaluate the SR and LR-trilinear fields on the same slices.\footnote{The slices are
selected with $|Q_{\rm HII}-Q_{\rm HII}^{*}|\le0.025$ around the HR slice nearest each stage $Q_{\rm
HII}^{*}$, with at least four slices per stage. We also use the same slices for the WST analysis.}
Only $477$, $439$ and $345$ of the $600$ HR histories provide such slices between $z=20$ and $5$.
Figure~\ref{fig:voxel} shows the percentage of voxels per bin, pooled over the lightcones that reach
each stage, for SR and for SR unnormalized, the decoder output $\tilde{x}_{\rm HII}$ before the
normalization. We measure the agreement by the overlap of two distributions, the percentage of
voxels they share when compared bin by bin, with a bin width of $0.1$ in $x_{\rm HII}$. SR overlaps
HR by $93$, $92$ and $93\%$ at the three stages, against $37$, $29$ and $36\%$ for LR-trilinear,
which concentrates its voxels in a single peak near the mean ionized fraction of the stage. The lower panels of Figure~\ref{fig:voxel} show the excess or deficit of voxels in each bin
relative to HR, $p_X-p_{\rm HR}$. At higher ionized fractions a larger percentage of HR voxels saturates at $x_{\rm HII}=1$, so at
$Q_{\rm HII}=0.5$ and $0.75$ most of the remaining difference lies near this bound. SR unnormalized
is smoother, and its percentage of voxels rises gradually toward $x_{\rm HII}=1$ and continues
beyond it. Its fraction below $0$ is larger earlier in reionization, and its fraction above $1$
grows as $Q_{\rm HII}$ approaches $1$. SR unnormalized overlaps HR by $85$, $78$ and $80\%$, and the
normalization raises the overlap by returning these voxels to the bounds.

\paragraph{Non-Gaussian morphology.}
\label{sec:results:wst}
We test the non-Gaussian information with the two-dimensional wavelet scattering transform
(WST)~\citep{Mallat2012Scattering, Greig2022WST, Andreux2020Kymatio}, applied to the transverse slices of the
three matched ionization stages. The transform convolves each slice with Morlet wavelets $\psi_j$ of
central wavenumber $k_j\simeq1.2\times2^{-j}\,h\,{\rm Mpc}^{-1}$ at $L=8$ orientations. For a slice
$I$, $S_0=\langle I\rangle$ is its mean, $S_1(j)=\langle|I\star\psi_j|\rangle$ the mean modulus of
the slice convolved with $\psi_j$, and
$S_2(j_1,j_2)=\langle||I\star\psi_{j_1}|\star\psi_{j_2}|\rangle$ the mean modulus of that modulus
field convolved again at a larger scale $j_2>j_1$, each averaged over the $L$ orientations. The
reduced coefficients $s_1(j)=S_1(j)/S_0$ measure the amount of structure at scale $j$ relative to
the mean, and $s_2(j_1,j_2)=S_2(j_1,j_2)/S_1(j_1)$ measures how unevenly the structure at scale
$j_1$ is spread on the larger scale $j_2$. This is interpreted as coupling between scales. We use
$J=5$ scales, whose central wavenumbers span the bands of the Fourier analysis.
$k_0\simeq1.2\,h\,{\rm Mpc}^{-1}$ lies beyond half the HR Nyquist frequency, $k_1$ and $k_2$ lie
between it and the LR Nyquist frequency, where the model generates the modes, and $k_3$ and $k_4$
lie below the LR Nyquist frequency, where the LR input resolves them.

\begin{figure}
\centering
\includegraphics[width=\columnwidth]{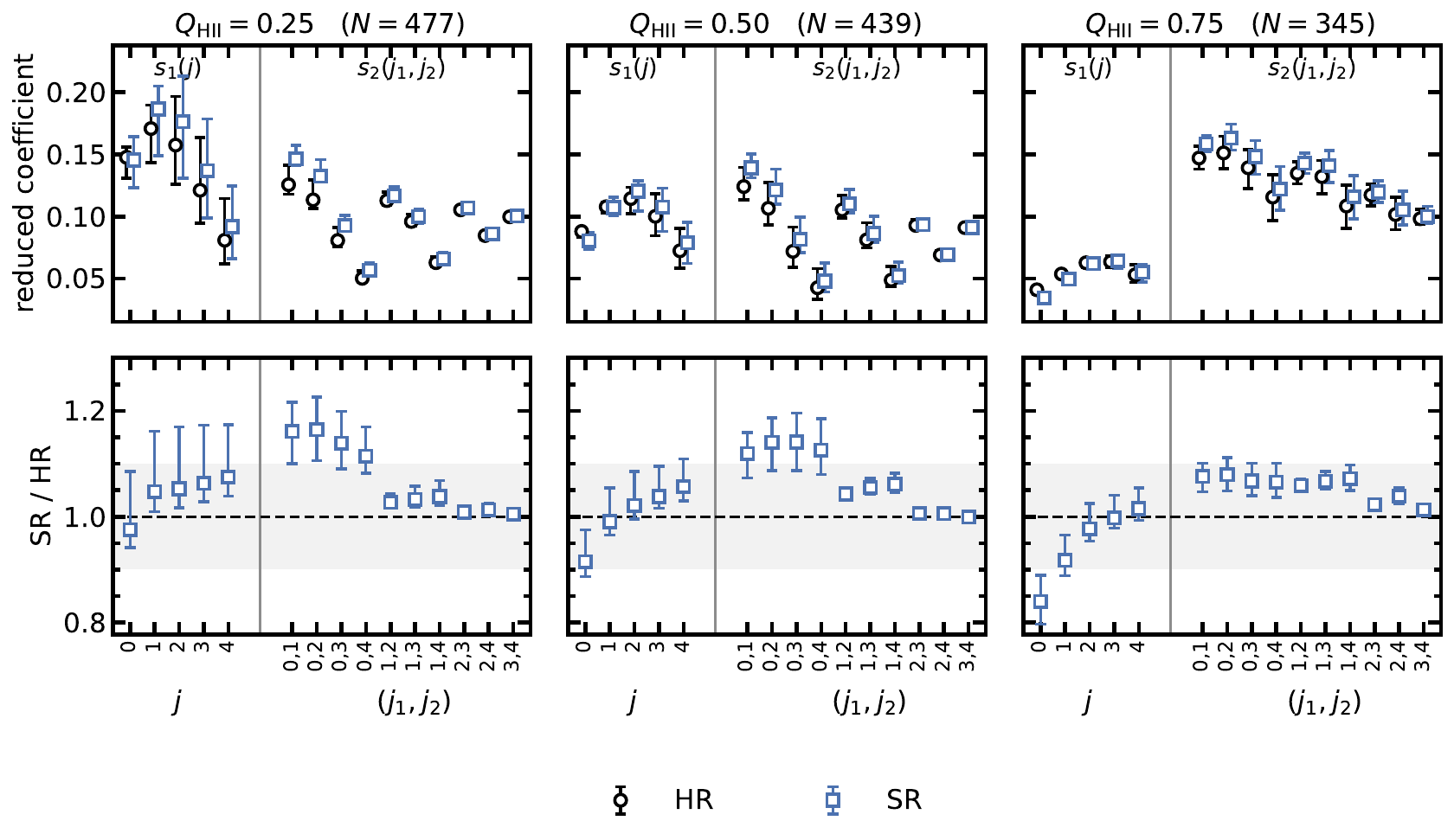}
\caption{Reduced wavelet scattering coefficients of HR and SR at three matched ionization stages
(top) and their ratio SR/HR (bottom, shaded band $\pm10\%$), for $s_1(j)$ and $s_2(j_1,j_2)$
coefficients at central wavenumbers from $k_0\simeq1.2$ ($j=0$) to $k_4\simeq0.07\,h\,{\rm Mpc}^{-1}$
($j=4$). Each wavelet corresponds to a band of wavenumbers around $k_j$. The $j=0$ band lies beyond
half the HR Nyquist frequency, where SR retains only part of the HR power, so $S_1(0)$ falls below
HR. The reduced $s_2(0,j_2)=S_2(0,j_2)/S_1(0)$ divides by this smaller $S_1(0)$ and therefore rises
above HR. The square markers are medians over lightcones and error bars the $1\sigma$ range.}
\label{fig:wst}
\end{figure}

Figure~\ref{fig:wst} compares the reduced coefficients of SR and HR at the three matched
stages, and Table~\ref{tab:wst} lists the median ratio SR/HR of $s_1(j)$ at
each scale and of $s_2(j_1,j_2)$ averaged over $j_2>j_1$. SR agrees with HR to within $10\%$ at every scale except the finest, $j=0$,
where the deviation reaches $15\%$. This scale, $k_0\simeq1.2\,h\,{\rm Mpc}^{-1}$, lies beyond half
the HR Nyquist frequency, where the reconstruction retains only part of the HR power
(\S\ref{sec:results:pk}). The missing power lowers $S_1(0)$ of SR below HR, and since
$s_2(0,j_2)$ divides $S_2(0,j_2)$ by $S_1(0)$, the same loss raises the second-order ratio above HR. The $j=1$ wavelet spans $0.27$-$0.96\,h\,{\rm Mpc}^{-1}$
at half maximum. SR thus reproduces both the amount of structure
and its coupling between scales at this and all larger scales.

\begin{table}
\centering
\small
\caption{Ratio SR/HR of the reduced wavelet scattering coefficients at three matched ionization
stages, as medians over the $N=477$, $439$ and $345$ lightcones that reach each stage. $k_j$ is the
central wavenumber of the first wavelet, and the second-order ratios are averaged over the coarser
second scale $j_2>j_1$. The $1\sigma$ scatter over lightcones is at most $0.08$.}
\label{tab:wst}
\begin{tabular}{l c ccc}
\hline
 & & \multicolumn{3}{c}{SR\,/\,HR} \\
coefficient & $k_j$ [$h$\,Mpc$^{-1}$] & $Q_{\rm HII}=0.25$ & $Q_{\rm HII}=0.50$ & $Q_{\rm HII}=0.75$ \\
\hline
$s_1(j=0)$ & $1.2$ & $0.98$ & $0.91$ & $0.84$ \\
$s_1(j=1)$ & $0.59$ & $1.05$ & $0.99$ & $0.92$ \\
$s_1(j=2)$ & $0.29$ & $1.05$ & $1.02$ & $0.98$ \\
$s_1(j=3)$ & $0.15$ & $1.06$ & $1.04$ & $1.00$ \\
$s_1(j=4)$ & $0.07$ & $1.07$ & $1.06$ & $1.02$ \\
\hline
$s_2(j_1=0)$ & $1.2$ & $1.15$ & $1.13$ & $1.07$ \\
$s_2(j_1=1)$ & $0.59$ & $1.03$ & $1.05$ & $1.07$ \\
$s_2(j_1=2)$ & $0.29$ & $1.01$ & $1.01$ & $1.03$ \\
$s_2(j_1=3)$ & $0.15$ & $1.00$ & $1.00$ & $1.01$ \\
\hline
\end{tabular}
\end{table}

\subsection{Population-level recovery: spectral statistics}
\label{sec:results:pk}

\begin{figure}
\centering
\includegraphics[width=\columnwidth]{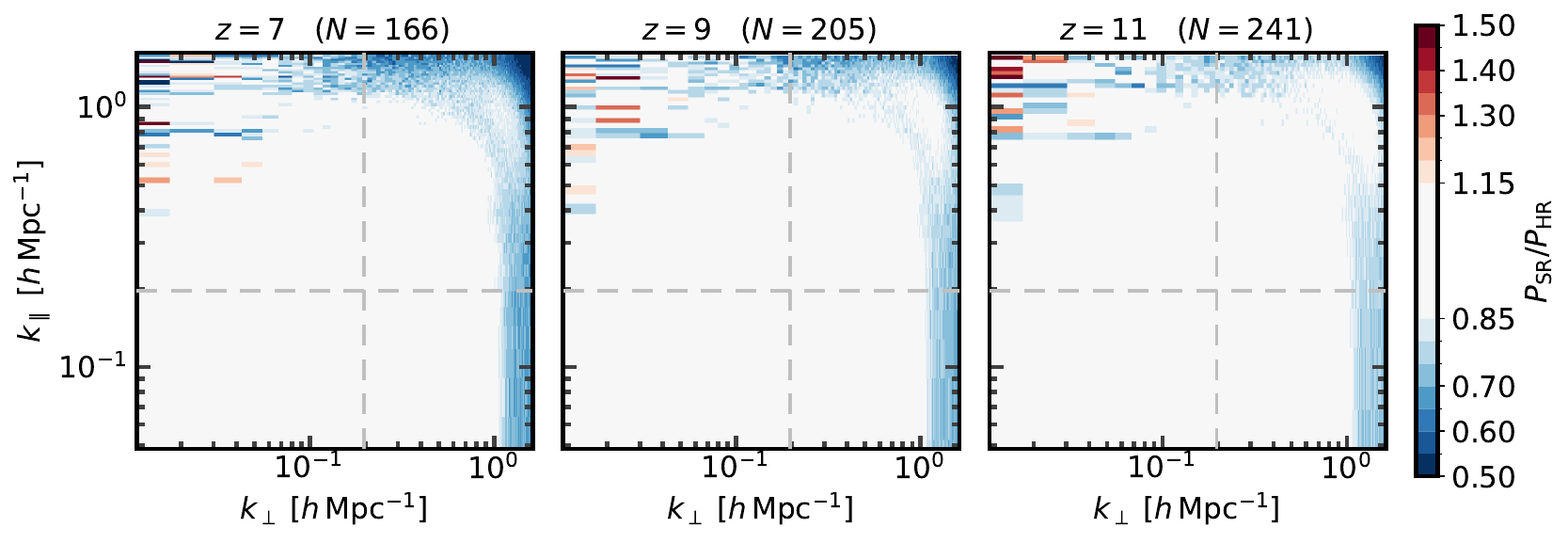}
\includegraphics[width=\columnwidth]{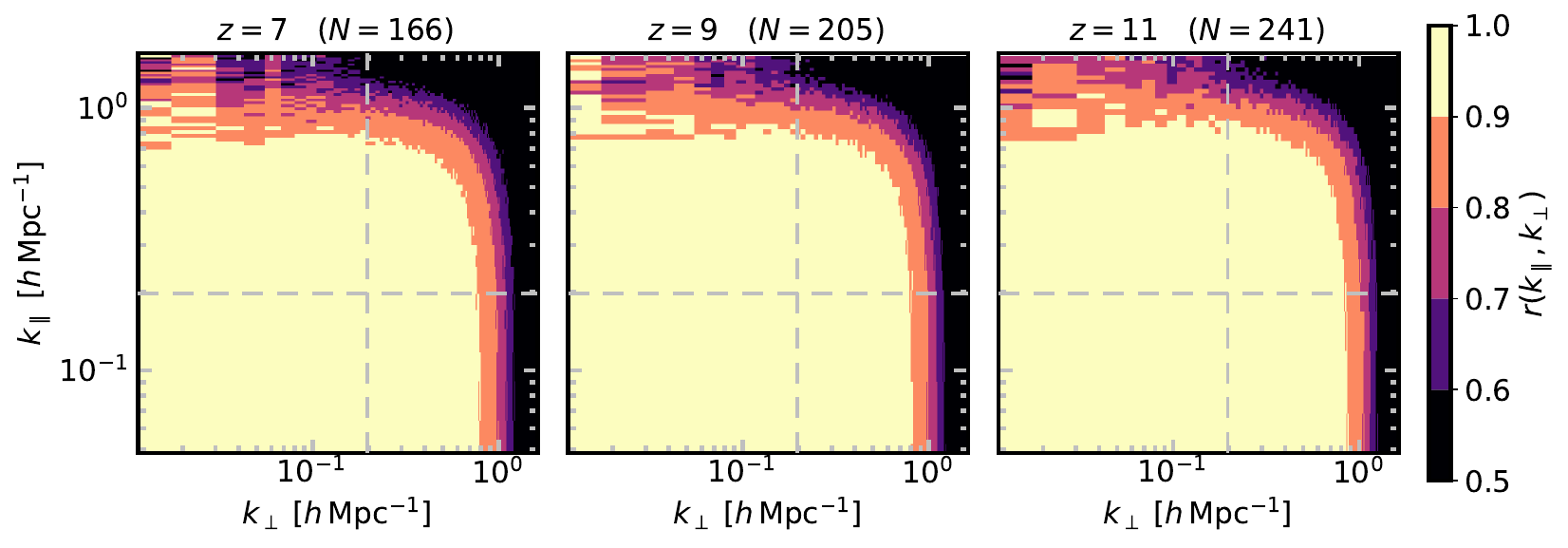}
\caption{Population-median cylindrical power ratio $P_{\rm SR}/P_{\rm HR}$ (top) and cross-coherence
$r(k_\parallel,k_\perp)$ (bottom) in the $(k_\parallel,k_\perp)$ plane at $z=7$, $9$ and $11$, over
the $N$ lightcones partially ionized in each $w_z=1$ window. White cells mark $0.85\le P_{\rm
SR}/P_{\rm HR}\le1.15$ and dashed lines the LR Nyquist frequency.}
\label{fig:psdev}
\end{figure}

We now extend the Fourier comparison of \S\ref{sec:results:qual} to the test cohort. Figure~\ref{fig:psdev} shows the population-median power ratio $P_{\rm{SR}}/P_{\rm{HR}}$ and coherence $r$ over the $(k_\parallel,k_\perp)$ plane on $w_z=1$ windows at $z=7$, $9$ and $11$, over the
$166$, $205$ and $241$ lightcones that are partially ionized there ($0.05<Q_{\rm HII}<0.95$).

Within half the HR Nyquist frequency SR recovers $0.97^{+0.10}_{-0.09}$ of the HR power at a
coherence of $0.96^{+0.02}_{-0.14}$. Below $k_{\rm Nyq}^{\rm LR}$, where the LR input resolves the
modes, SR keeps about $1.01$ of the power at $r>0.99$ and beyond $k_{\rm Nyq}^{\rm HR}/2$ about
$0.82$ at $r\simeq0.5$. These medians are weighted toward early reionization, since $226$ of the $612$ windows have $Q_{\rm
HII}\le0.2$. Sorted by ionization state, the windows between $k_{\rm Nyq}^{\rm LR}$ and $k_{\rm
Nyq}^{\rm HR}/2$ fall from a power ratio of $1.00$ at $0.2<Q_{\rm HII}\le0.5$ to $0.81$ at
$0.8<Q_{\rm HII}<0.95$, and from a coherence of $0.97$ to $0.74$ between the most neutral and the
most ionized windows, while below $k_{\rm Nyq}^{\rm LR}$ the coherence stays at $r\ge0.99$. The loss
is therefore confined to the scales the model must generate. 

The kSZ and $21$-cm signals trace the density-weighted fields $x_{\rm HII}(1+\delta)$ and
$x_{\rm HI}(1+\delta)$. Appendix~\ref{app:density} shows that SR preserves the structure of both
fields in the same redshift windows. At the scales the model generates, the suppression of power in the ionized field $x_{\rm HII}(1+\delta)$
is accompanied by a boost of power in the neutral field $x_{\rm HI}(1+\delta)$, which stays within $15\%$ of HR up to half
the HR Nyquist frequency and exceeds HR beyond.

\begin{figure}
\centering
\includegraphics[width=\columnwidth]{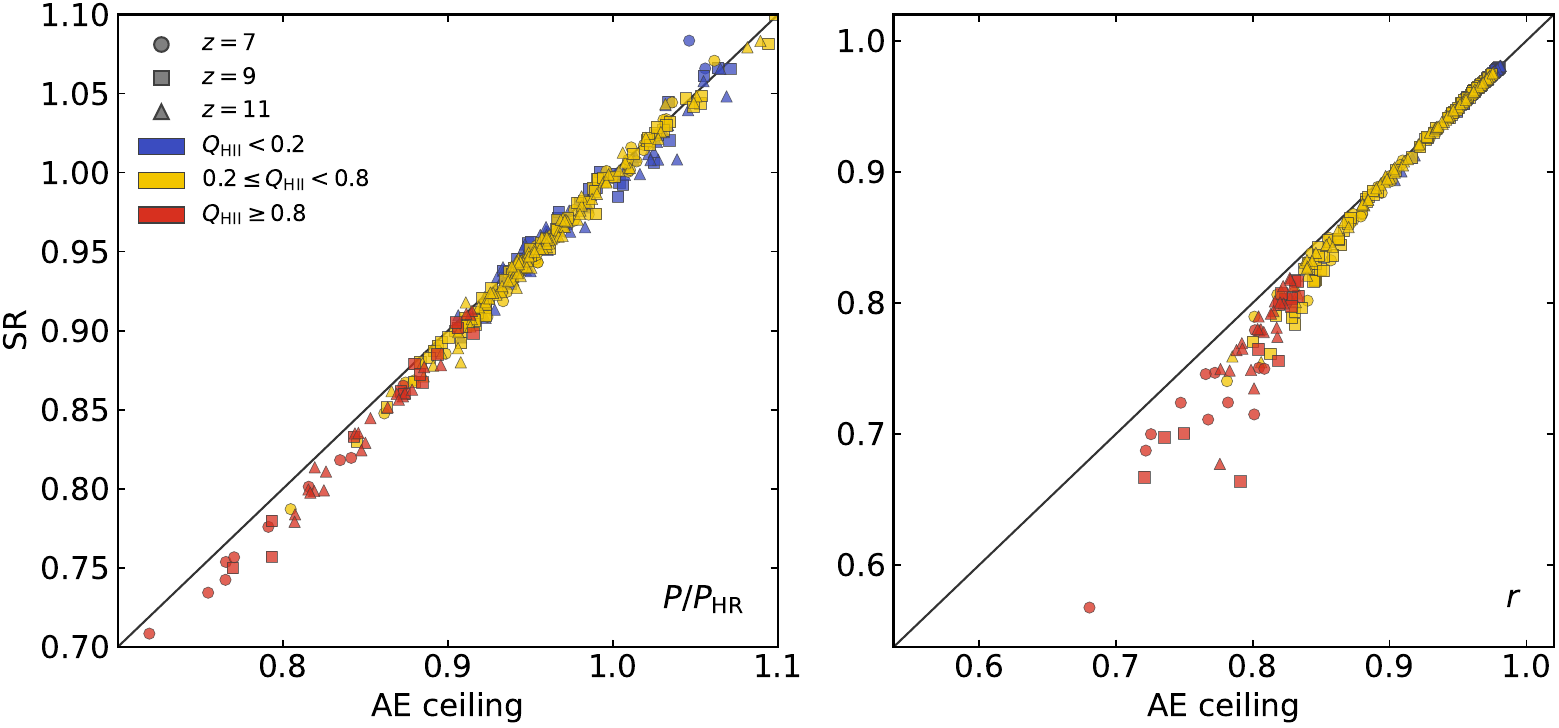}
\caption{Per-redshift-window medians of the cylindrical power ratio (left) and coherence (right) within half
the HR Nyquist frequency, SR against the autoencoder ceiling, the HR field encoded and decoded by
the frozen autoencoder, on the same $w_z=1$ windows at $z=7$ (circles), $9$ (squares) and $11$
(triangles). Colours mark the volume-averaged ionized fraction of the window.}
\label{fig:ceiling}
\end{figure}

\paragraph{Autoencoder ceiling.}
\label{sec:results:ceiling}
The autoencoder ceiling refers to the HR field encoded and decoded by the frozen autoencoder. Since
the denoiser is trained to recover the encoded HR latent, this field is the target of the
reconstruction. Figure~\ref{fig:ceiling} compares SR with the ceiling redshift window by window. In every window, the power ratio and coherence of SR
match those of the ceiling, so the denoiser recovers $\sim100\%$ of what the latent retains. SR falls
outside 15$\%$ of the HR power within half the HR Nyquist frequency only in the windows with $Q_{\rm HII}>0.8$, where the ceiling shows the same deficit. Near
the end of reionization the neutral hydrogen occupies small, isolated islands, and encoding and
decoding the HR field smooths them and suppresses their contribution to the power spectrum. The
compression of the latent therefore limits the attainable fidelity, while it keeps every step of the
reverse chain inexpensive. The chain evaluates the full denoiser on every latent site at
each of its $15$ steps, so its cost grows with the size of the latent. A less compressive latent
would likely raise this ceiling and make the autoencoder cheaper, but would make every diffusion step more expensive, eightfold for a latent
doubled along each axis. We therefore keep the compact latent for its lower sampling cost.

\subsection{Dependence on astrophysical parameters}
\label{sec:results:cond}

\begin{figure}
\centering
\includegraphics[width=\columnwidth]{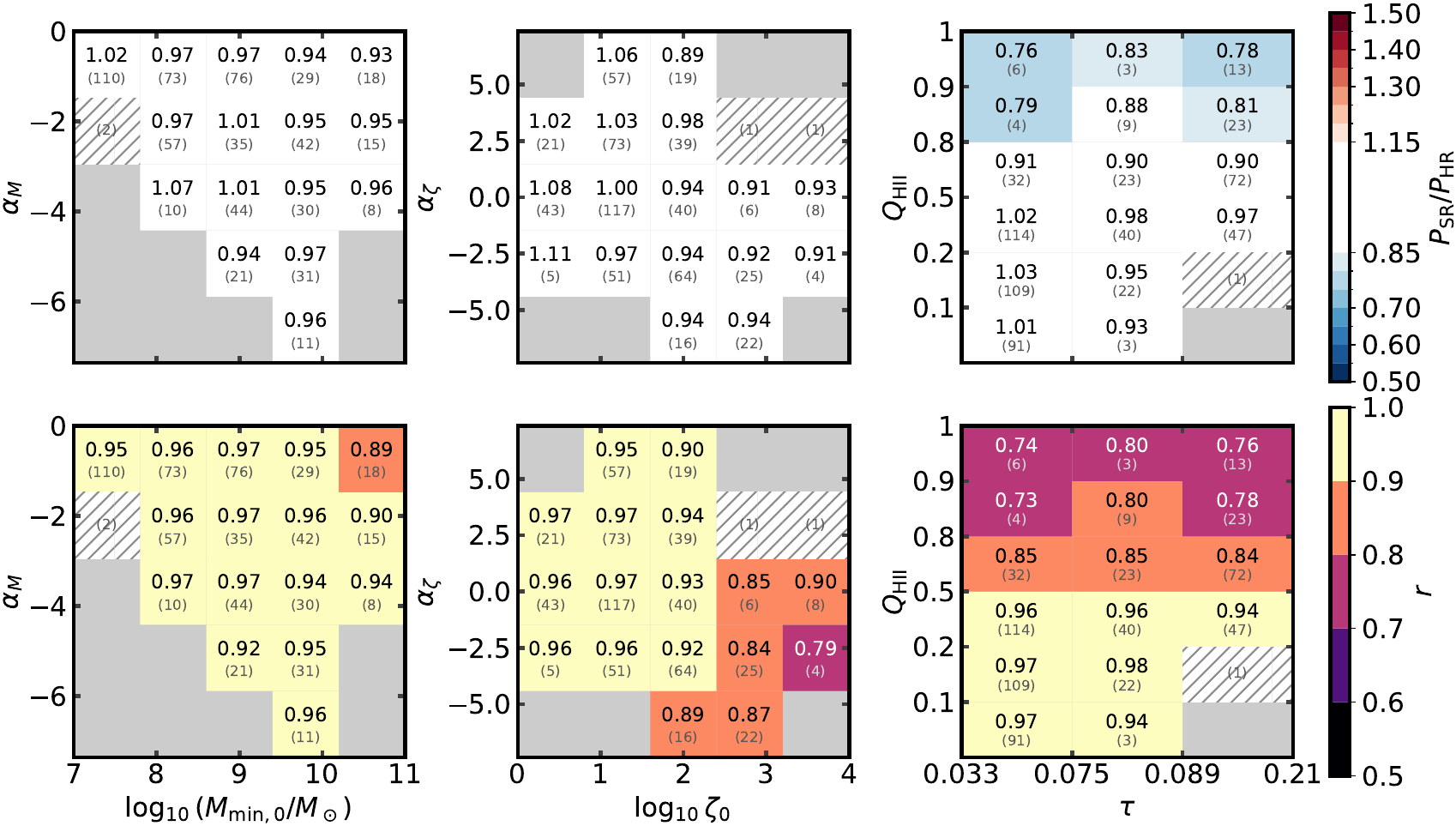}
\caption{Median of the per-redshift-window cylindrical power ratio $P_{\rm SR}/P_{\rm HR}$ (top) and
cross-coherence $r$ (bottom) within half the HR Nyquist frequency, for the $w_z=1$ windows at
$z=7$, $9$ and $11$ with $0.05<Q_{\rm HII}<0.95$, in bins of $(\log_{10}M_{\rm min,0},\alpha_M)$
(left), $(\log_{10}\zeta_0,\alpha_\zeta)$ (middle), and the optical depth $\tau$ of the lightcone
and the volume-averaged ionized fraction $Q_{\rm HII}$ of the window (right). The number of windows
in each bin is given in parentheses. The $\tau$ bin edges, $\tau = 0.075$ and 0.089, lie $3\sigma$ and $5\sigma$ above the
Planck best fit $\tau=0.054\pm0.007$~\citep{Planck2020}. White cells mark
$0.85\le P_{\rm SR}/P_{\rm HR}\le1.15$, hatched cells contain fewer than three windows and grey
cells contain none.}
\label{fig:cond}
\end{figure}

We test whether recovery depends on the source model by binning the per-window median power
ratio and coherence within half the HR Nyquist frequency over the prior of the conditioning vector
(\S\ref{sec:data:dataset}), on the windows of \S\ref{sec:results:pk}, the $w_z=1$ windows at
$z=7$, $9$ and $11$ of the lightcones partially ionized there. Figure~\ref{fig:cond} shows the median over the windows in
each cell of the $(\log_{10}M_{\rm min,0},\alpha_M)$ and $(\log_{10}\zeta_0,\alpha_\zeta)$ planes
(left and middle columns), for the power ratio (top) and the coherence (bottom). Since every source
parameter enters the denoiser as conditioning, a poorly learned region of the prior would show as
low recovery in the cells it covers.

The recovery varies little across the prior. Most cell medians in these planes lie within
$15\%$ of HR. The coherence is lower only in the cells of high $\zeta_0$, which hold the most
ionized windows. A lightcone with a higher optical depth reionizes earlier, so at a given redshift
its window is more ionized. The right column of Figure~\ref{fig:cond} shows the recovery binned in
both $\tau$ and the ionized fraction of the window. The recovery falls only in the highly ionized
windows, where the latent representation smooths the remaining neutral islands
(\S\ref{sec:results:ceiling}). The optical depth integrates the ionization history, so reionization at higher redshifts increases $\tau$. In the $\tau$ bins, placed $3\sigma$ and $5\sigma$ above
the Planck best fit $\tau=0.054\pm0.007$~\citep{Planck2020}, the lightcones consistent with the
measurement and those far above it recover the same power and coherence at a given ionized
fraction. The recovery is therefore set by the ionized fraction of the window.

\subsection{Computational cost}
\label{sec:results:cost}

Table~\ref{tab:cost} lists the inference costs of each stage across $600$ test lightcones on each GPU and over $40$ test lightcones on a single CPU core. On NVIDIA A100, a reconstruction takes $4.04$\,s per lightcone. This includes $2.47$\,s for encoding, $0.765$\,s for the 15-step diffusion chain, and $0.804$\,s for decoding. A single CPU core takes $1390$\,s, with $793$\,s for encoding, $216.8$\,s for the chain, and $381$\,s for decoding. On both devices the autoencoder accounts for over $80\%$ of the time, because it operates at the HR resolution. 

\begin{table}
\centering
\small
\setlength{\tabcolsep}{4pt}
\caption{Per-lightcone inference cost over the $N$ test
lightcones on each device. ``Encode'' stage includes the trilinear upsampling of the LR lightcone, and
``diffusion'' stage is the $15$-step reverse chain.}
\label{tab:cost}
\begin{tabular}{l c cccc}
\hline
Device & $N$ & encode\,[s] & diffusion\,[s] & decode\,[s] & total\,[s] \\
\hline
A100 (80\,GB) & 600 & $2.47 \pm 0.04$ & $0.765 \pm 0.004$ & $0.804 \pm 0.002$ & $4.04 \pm 0.04$ \\
A30 (24\,GB) & 600 & $5.96 \pm 0.06$ & $1.33 \pm 0.02$ & $1.79 \pm 0.03$ & $9.09 \pm 0.04$ \\
CPU, 1 core & 40 & $793 \pm 1$ & $216.8 \pm 0.5$ & $381 \pm 1$ & $1390 \pm 2$ \\
\hline
\end{tabular}
\end{table}

The HR \texttt{SCRIPT} run takes a median $7625$\,s per lightcone on a single CPU core from the cached matter fields, while the LR run takes $9.5$\,s. Reconstruction on the A100 is thus approximately $560$ times faster, including the LR run. Even on a single CPU core, generating the LR lightcone and super-resolving it is $5.4$ times faster. An LR lightcone also holds $512$ times fewer voxels than its
HR counterpart, so LR lightcones can be generated and stored in bulk and super-resolved only where
an analysis needs them.

\subsection{Super-resolution at \texorpdfstring{$16\times$}{16x}}
\label{sec:results:x16}

At $16\times$, the LR lightcones, constructed from \texttt{SCRIPT} evaluations of the
ionization field on $16^3$ grids, have shape $64\times16\times16$ for the same
$1024\times256\times256$ HR targets. Since the autoencoder was trained on HR fields alone and the LR
input is upsampled to the HR grid before encoding, the autoencoder is unchanged and only the
denoiser needs finetuning. The denoiser is finetuned from the $8\times$ weights on the same
training set, now mapping the $16\times$ LR latents $\mathbf{z}_{\rm LR}$ to the same HR latent
targets $\mathbf{z}_0$.

Table~\ref{tab:x16pop} compares the $8\times$ denoiser applied directly to the $16\times$ LR
lightcones (zero-shot), the finetuned denoiser, the $8\times$ model and the autoencoder
ceiling. Zero-shot, the power ratio within half the HR Nyquist frequency falls from
$0.90$--$1.01$ to $0.69$--$0.85$, and the coherence drops below its $8\times$ value by $0.07$ at
$Q_{\rm HII}=0.25$ and by $0.30$ at $0.75$. Finetuning removes this deficit, and the
denoiser then matches the autoencoder ceiling and the $8\times$ model at every stage. On
the training realization, the latent representation therefore sets the fidelity ceiling
up to $16\times$, independent of the LR resolution. For reference, Appendix~\ref{app:x16} presents the
cylindrical spectra of the $\tau=0.054$ and $\tau=0.102$ lightcones at $16\times$.

\begin{table}
\centering
\small
\setlength{\tabcolsep}{4pt}
\caption{Recovery at $16\times$ by the $8\times$ denoiser applied without finetuning (zero-shot)
and by the denoiser finetuned for the $16\times$ input (\ft{SR}), against the $8\times$ model and
the autoencoder ceiling, the HR field encoded and decoded by the frozen autoencoder. $R_P$ is the
median cylindrical power ratio $P_{\rm SR}/P_{\rm HR}$ and $r$ the median SR--HR cross-coherence
within half the HR Nyquist frequency, given as the median and $1\sigma$ range over the $452$, $445$
and $401$ test lightcones that reach each stage.}
\label{tab:x16pop}
\begin{tabular}{l l cc cc cc}
\hline
 & & \multicolumn{2}{c}{$Q_{\rm HII}=0.25$} & \multicolumn{2}{c}{$Q_{\rm HII}=0.50$} & \multicolumn{2}{c}{$Q_{\rm HII}=0.75$} \\
model & input & $R_P$ & $r$ & $R_P$ & $r$ & $R_P$ & $r$ \\
\hline
SR, zero-shot & $16\times$ & $0.85^{+0.06}_{-0.07}$ & $0.89^{+0.04}_{-0.09}$ & $0.85^{+0.09}_{-0.08}$ & $0.72^{+0.11}_{-0.16}$ & $0.69^{+0.07}_{-0.16}$ & $0.51^{+0.11}_{-0.18}$ \\
\ft{SR} & $16\times$ & $1.01^{+0.07}_{-0.06}$ & $0.96^{+0.01}_{-0.02}$ & $0.97^{+0.08}_{-0.05}$ & $0.90^{+0.03}_{-0.04}$ & $0.90^{+0.07}_{-0.05}$ & $0.81^{+0.02}_{-0.04}$ \\
SR & $8\times$ & $1.01^{+0.07}_{-0.05}$ & $0.96^{+0.01}_{-0.03}$ & $0.97^{+0.08}_{-0.05}$ & $0.90^{+0.03}_{-0.04}$ & $0.90^{+0.07}_{-0.05}$ & $0.81^{+0.02}_{-0.04}$ \\
AE ceiling & HR & $1.02^{+0.07}_{-0.06}$ & $0.96^{+0.01}_{-0.02}$ & $0.98^{+0.08}_{-0.05}$ & $0.90^{+0.03}_{-0.03}$ & $0.90^{+0.07}_{-0.05}$ & $0.83^{+0.02}_{-0.02}$ \\
\hline
\end{tabular}
\end{table}

\subsection{Generalization to a new matter realization}
\label{sec:results:newseed}

\begin{figure}
\centering
\includegraphics[width=0.97\columnwidth]{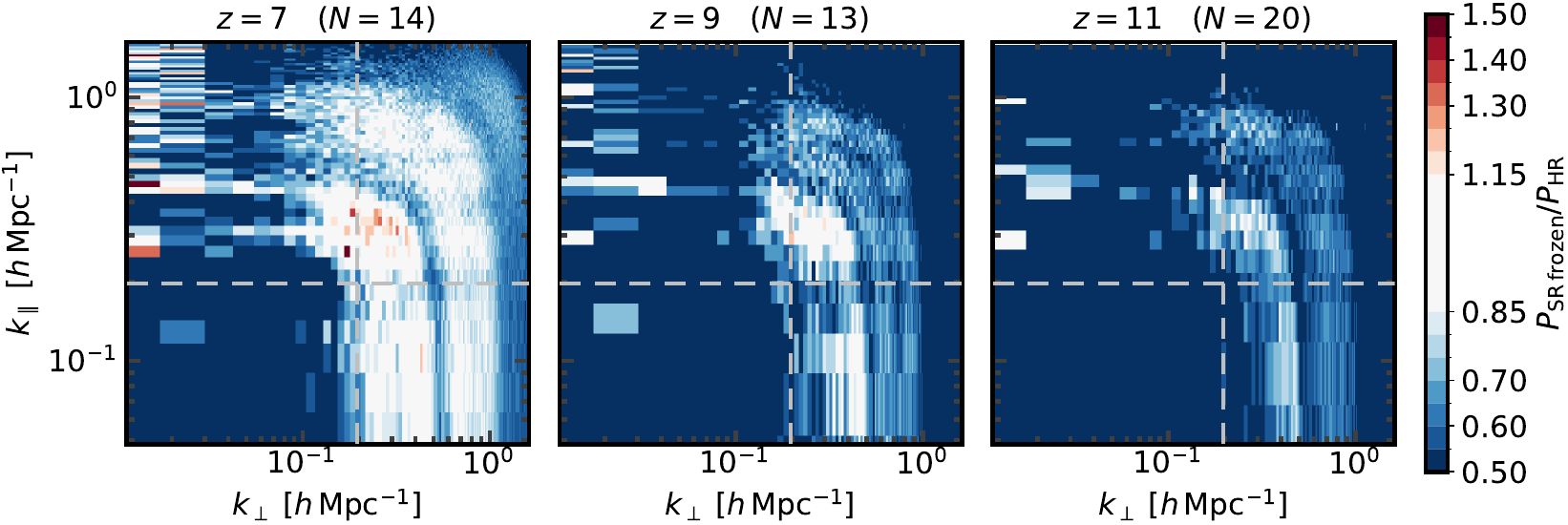}
\includegraphics[width=0.97\columnwidth]{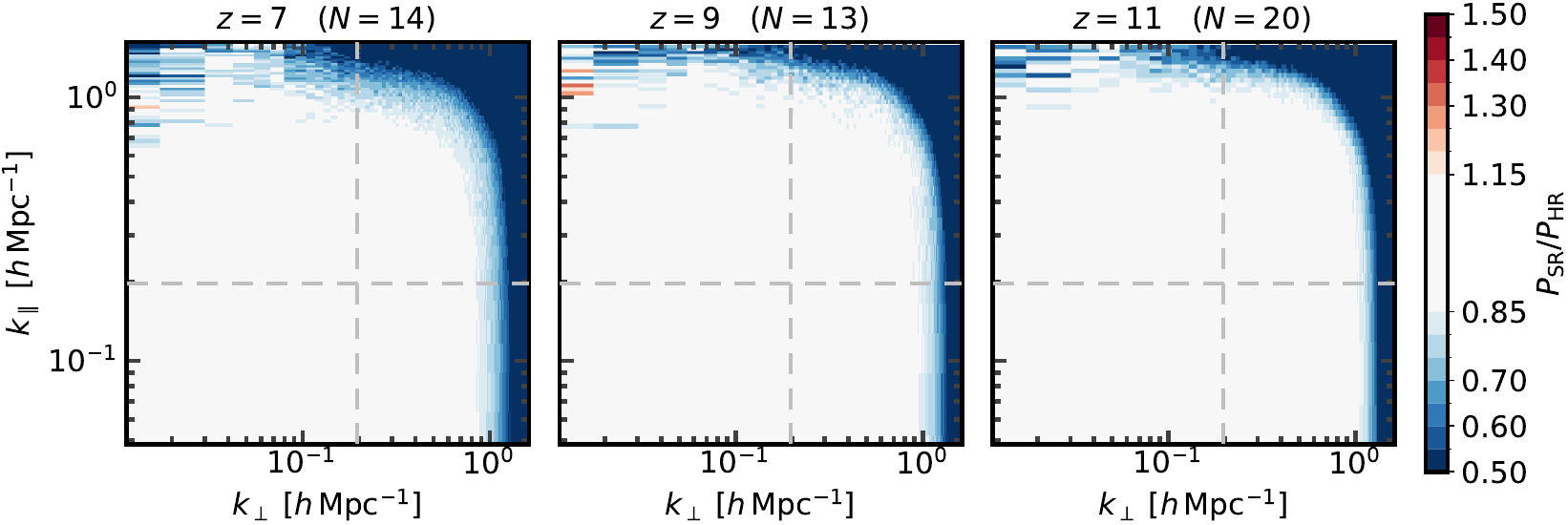}
\includegraphics[width=0.97\columnwidth]{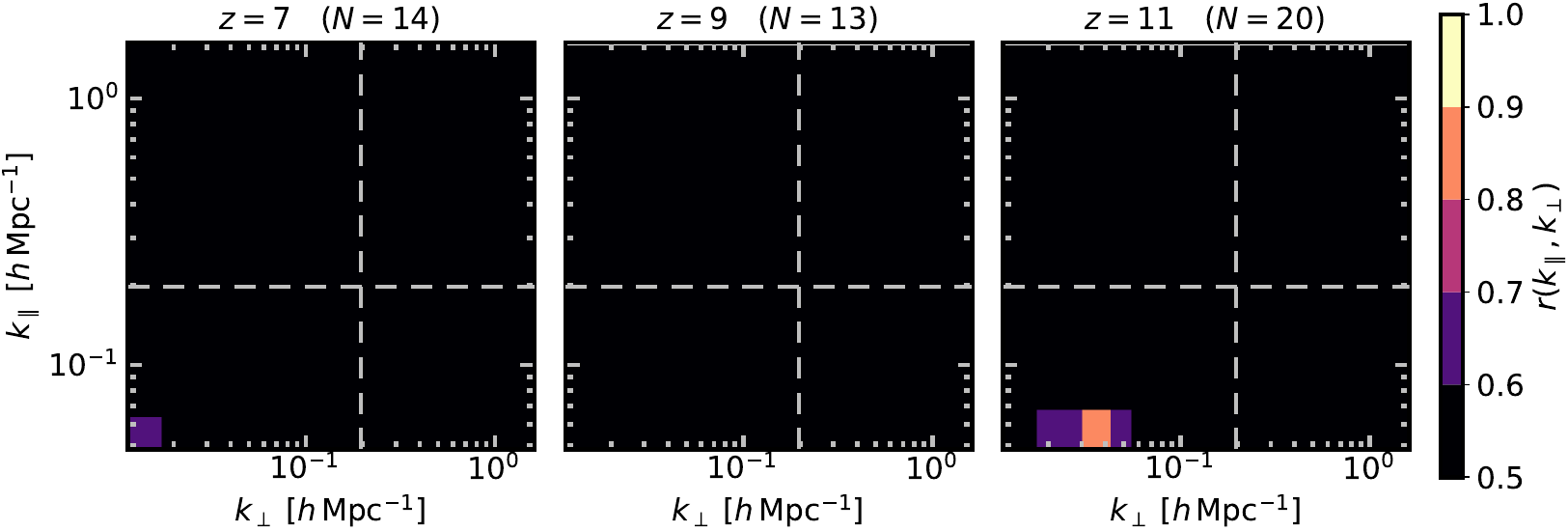}
\includegraphics[width=0.97\columnwidth]{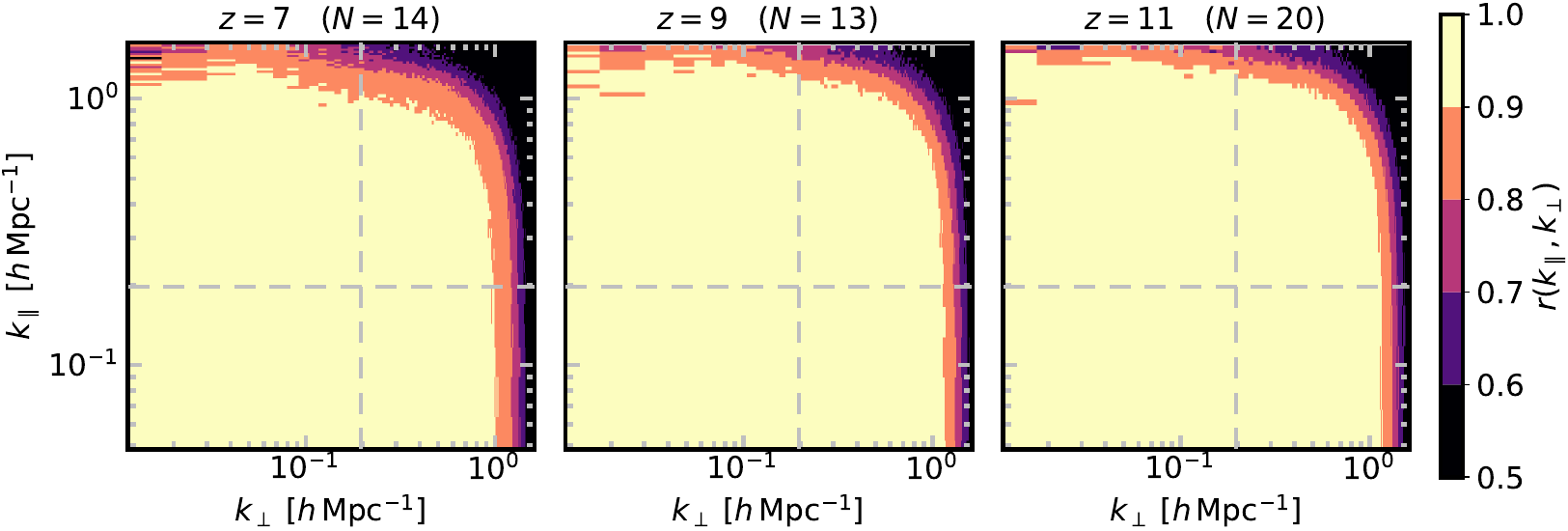}
\caption{Population-median cylindrical power ratio $P_{\rm SR}/P_{\rm HR}$ (top two rows) and
cross-coherence $r(k_\parallel,k_\perp)$ (bottom two rows) on the new matter realization at $z=7$,
$9$ and $11$, for the frozen and then the finetuned model, over the $N$ lightcones partially ionized
in each window. White cells mark $0.85\le P_{\rm SR}/P_{\rm HR}\le1.15$ and dashed lines the LR
Nyquist frequency.}
\label{fig:newseed:popratio}
\end{figure}

Evaluated on $53$ test lightcones from a new realization of the matter density field, the
frozen \SuperSCRIPT{} model fails to reconstruct the small scales. Table~\ref{tab:newseed} lists the power ratio and the coherence at the three ionization stages. The coherence within half the HR
Nyquist frequency is at most $0.01$, against $0.81$-$0.96$ on the original realization.
Figure~\ref{fig:newseed:popratio} shows that SR loses the HR power and coherence even below
the LR Nyquist frequency. We find that both the autoencoder and the denoiser lose accuracy on the new realization. The ceiling of the frozen autoencoder degrades less than SR. It retains most of the power within half the HR Nyquist frequency, but at a reduced coherence. Since neither the LR ionization field nor the global conditioning vector supplies the unresolved density field that sets the unresolved ionization structure, the model likely learns this structure as a property of the training realization.

\begin{figure}
\centering
\includegraphics[width=\columnwidth]{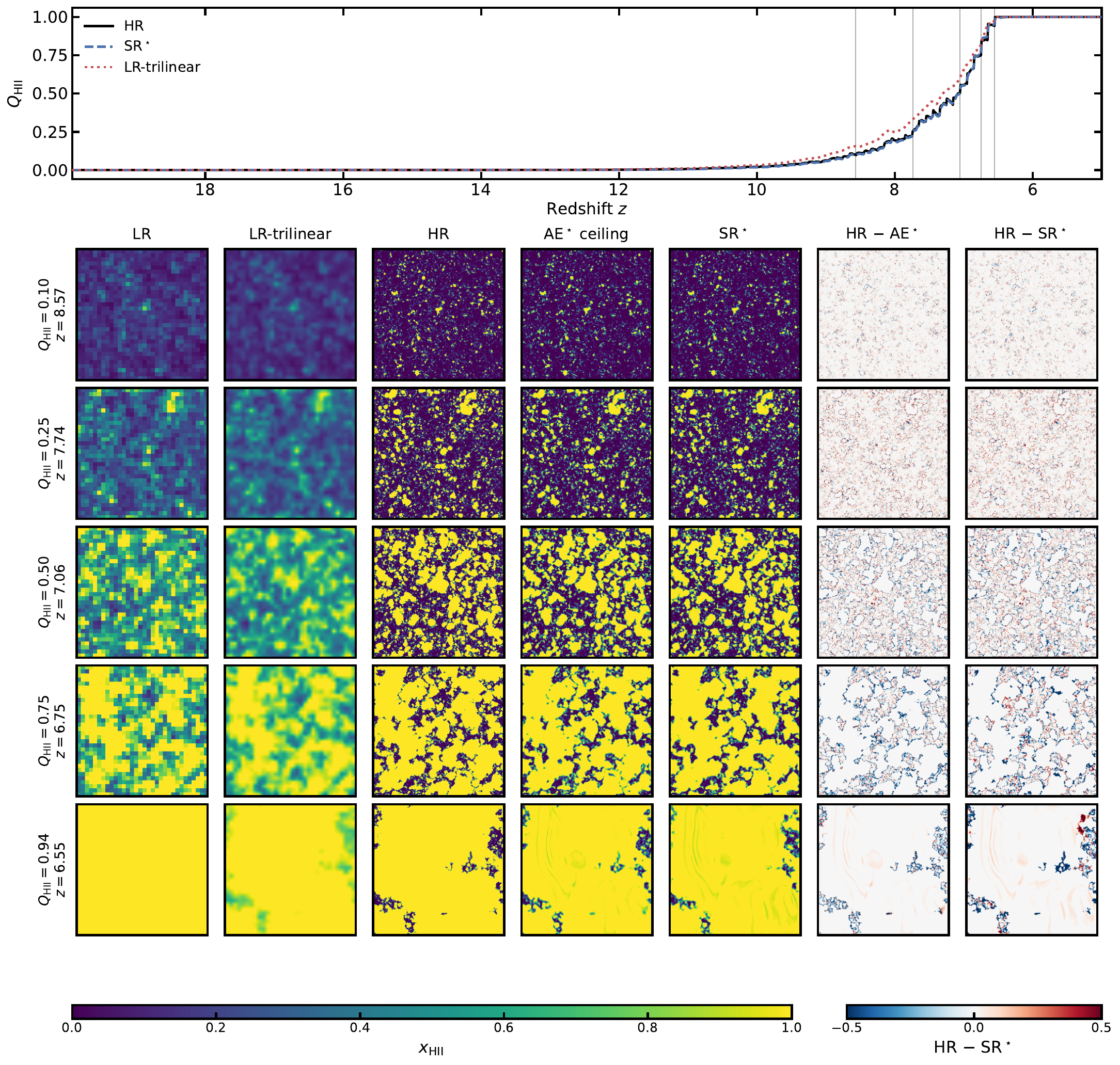}
\caption{Reconstruction by the finetuned model of the lightcone with optical depth $\tau=0.054$
on the new matter realization. The top panel shows the ionization histories $Q_{\rm HII}(z)$
of HR, SR and LR-trilinear, with vertical lines marking the redshifts of the slices. The bottom
panel shows transverse slices of LR, LR-trilinear, HR, the ceiling of the finetuned autoencoder
(\ft{AE}), the HR field encoded and decoded by it, and \ft{SR} at five stages of reionization,
followed by the residuals $x_{\rm HII}^{\rm HR-AE}$ and $x_{\rm HII}^{\rm HR-SR}$. \ft{SR} and
\ft{AE} are the models finetuned on the new realization.}
\label{fig:newseed:showcase}
\end{figure}

We therefore finetune both stages on $600$ additional lightcones of the new realization,
$10\%$ of the original training set, drawn over the same source-model prior. The autoencoder starts
from the frozen weights and trains for $100$ epochs on full volumes with the reconstruction loss. We encode the training set with the finetuned autoencoder and finetune the denoiser on these latents, warm-started from the $8\times$ weights. Finetuning restores the coherence at all three stages. SR closely matches the ceiling of the finetuned autoencoder in both power and coherence. Figure~\ref{fig:newseed:showcase} shows the finetuned reconstruction of the $\tau=0.054$ lightcone of \S\ref{sec:results:qual} on the new realization, and Appendix~\ref{app:newseed} presents the cylindrical spectra of the $\tau=0.054$ and $\tau=0.102$ lightcones for reference. On the original realization, however, the finetune degrades the reconstruction. For the four lightcones of \S\ref{sec:results:qual}, the coherence of the autoencoder ceiling above $k_{\rm Nyq}^{\rm LR}$ falls from $0.86$ to $0.51$. The finetuned model thus specializes to the new realization.

\begin{table}
\centering
\small
\setlength{\tabcolsep}{4pt}
\caption{Recovery on the new matter realization by the frozen and finetuned models and the
ceilings of the frozen and finetuned autoencoders, the HR field encoded and decoded by each, with
the frozen model on the original realization for reference. \ft{SR} and \ft{AE} are the models finetuned on the new
realization. $R_P$ is the median cylindrical power ratio $P_{\rm SR}/P_{\rm HR}$ and $r$ the median
SR--HR cross-coherence within half the HR Nyquist frequency, given as the median and $1\sigma$ range
over $37$, $42$ and $33$ lightcones of the new realization and $452$, $445$ and $401$ of the
original.}
\label{tab:newseed}
\begin{tabular}{l cc cc cc}
\hline
 & \multicolumn{2}{c}{$Q_{\rm HII}=0.25$} & \multicolumn{2}{c}{$Q_{\rm HII}=0.50$} & \multicolumn{2}{c}{$Q_{\rm HII}=0.75$} \\
 & $R_P$ & $r$ & $R_P$ & $r$ & $R_P$ & $r$ \\
\hline
\multicolumn{7}{l}{new realization} \\
SR & $0.42^{+0.25}_{-0.13}$ & $0.01^{+0.00}_{-0.00}$ & $0.81^{+0.16}_{-0.14}$ & $0.01^{+0.00}_{-0.00}$ & $0.88^{+0.15}_{-0.10}$ & $0.01^{+0.00}_{-0.00}$ \\
AE ceiling & $0.92^{+0.07}_{-0.05}$ & $0.22^{+0.10}_{-0.04}$ & $0.89^{+0.09}_{-0.03}$ & $0.30^{+0.11}_{-0.08}$ & $0.84^{+0.07}_{-0.03}$ & $0.43^{+0.08}_{-0.10}$ \\
\ft{SR} & $0.98^{+0.04}_{-0.03}$ & $0.99^{+0.00}_{-0.01}$ & $0.97^{+0.03}_{-0.04}$ & $0.97^{+0.01}_{-0.02}$ & $0.90^{+0.03}_{-0.03}$ & $0.93^{+0.02}_{-0.04}$ \\
\ft{AE} ceiling & $0.99^{+0.04}_{-0.03}$ & $0.99^{+0.00}_{-0.01}$ & $0.97^{+0.03}_{-0.03}$ & $0.98^{+0.01}_{-0.01}$ & $0.92^{+0.04}_{-0.04}$ & $0.95^{+0.01}_{-0.02}$ \\
\hline
\multicolumn{7}{l}{original realization} \\
SR & $1.01^{+0.07}_{-0.05}$ & $0.96^{+0.01}_{-0.03}$ & $0.97^{+0.08}_{-0.05}$ & $0.90^{+0.03}_{-0.04}$ & $0.90^{+0.07}_{-0.05}$ & $0.81^{+0.02}_{-0.04}$ \\
AE ceiling & $1.02^{+0.07}_{-0.06}$ & $0.96^{+0.01}_{-0.02}$ & $0.98^{+0.08}_{-0.05}$ & $0.90^{+0.03}_{-0.03}$ & $0.90^{+0.07}_{-0.05}$ & $0.83^{+0.02}_{-0.02}$ \\
\hline
\end{tabular}
\end{table}

\section{Discussion and conclusions}
\label{sec:discussion}
\label{sec:conclusions}

Semi-numerical simulations approximate radiative transfer through physically motivated
ionization prescriptions. This enables them to resolve ionization fields over cosmological volumes
at a cost that is compatible with inference~\citep{MesingerFurlanetto2007, Mesinger2011,
ChoudhuryParanjape2018}. Upcoming data make this trade-off more important, because inference from
them requires repeated forward simulations of the ionization field over large volumes. SKA-Low aims
for $21$-cm tomography~\citep{Koopmans2015}, while the Simons Observatory~\citep{SO2019} and future
surveys with CMB-S4-like sensitivity~\citep{CMBS42019} target high-significance kSZ
measurements~\citep{RaghunathanOmori2023,Jain2024kSZ}. These measurements, along with the patchy
optical depth, probe the same ionization field. The $21$-cm signal is observed as a lightcone, and
the kSZ signal and the patchy optical depth are integrated along the LOS, so maps of all three
require this field over the full lightcone. Therefore, with matched density and velocity fields,
ionization lightcone volume is the common ingredient of the reionization observables in both the
CMB and the $21$-cm signal.

Parameter inference from these signals has traditionally compressed each map into summary
statistics. These include the $21$-cm power spectrum~\citep{HERA2023} and kSZ power spectra or
trispectra~\citep{Jain2023framework, Jain2024kSZ, Raghunathan2024, Beringue2025ACT} for individual
probes, and the patchy optical depth-$21$-cm cross-correlation~\citep{Roy2020}, kSZ$^2$-$21$-cm
cross-spectra or the $21$-cm-kSZ-kSZ bispectrum~\citep{Ma2018,LaPlante2020} across probes. Beyond
these compressed statistics, Hotinli and Johnson~\cite{HotinliJohnson2022} demonstrated that kSZ
tomography can combine small-scale CMB temperature and $21$-cm measurements to reconstruct
large-scale velocity fields. The higher-order statistics and kSZ tomography couple modes from
few-Mpc ionized bubbles to modes on scales of several hundred comoving Mpc. Therefore, modelling
them necessitates simulations that resolve both scales, with the ionization field drawn from the
same realization as the density and velocity fields~\citep{Ma2018,LaPlante2020}.
Parameter-conditioned $21$-cm lightcone generators span independent realizations and are compared
with simulations only through ensemble statistics~\citep{Xia2026}. Super-resolution instead starts
off from the LR field of a given realization, retaining its large-scale structure, and generates
the unresolved morphology conditioned on it. The reconstructed ionization field is therefore
matched to the density and velocity fields of that realization.

We introduced \SuperSCRIPT, a conditional latent-diffusion model that generates such matched
lightcones at $8\times$ the resolution of its input. The LR lightcone provides the large-scale
ionization structure, while four source-model parameters and the optical depth condition the
unresolved morphology. A $15$-step residual-shift denoiser reconstructs the HR field within the
latent space of a frozen anisotropic autoencoder. Since both the LR input and the HR target come
from \texttt{SCRIPT}, the large-scale information converges across the two resolutions and the
small-scale ionization structure is learnt from the HR training target. At the midpoint of
reionization, where the ionization field is patchiest, the SR-to-HR power ratio and the
cross-coherence over the held-out test lightcones have medians of $0.97$ and $0.90$ within half the
HR Nyquist frequency (Table~\ref{tab:cyl:showcase}). The reconstruction's wavelet scattering
coefficients agree with HR to within $10\%$ on all except the finest scale. The reconstruction also
recovers the density-weighted fields that enter the kSZ and $21$-cm signals,
$x_{\rm HII}(1+\delta)$ at a coherence of $0.98$ and $x_{\rm HI}(1+\delta)$ at $0.91$--$0.93$
(Appendix~\ref{app:density}), and its optical depth is consistent with that of HR, well within the
$1\sigma$ uncertainty of the Planck measurement (\S\ref{sec:results:pdf}). The mode-by-mode match
of the reconstruction with its own HR lightcone therefore enables cross-correlations between the
probes. The training set draws the source-model parameters from a broad prior, yielding
reionization histories with optical depth spanning from $\tau=0.030$ to $0.211$
(\S\ref{sec:data:dataset}). We investigated how the reconstruction fidelity varies over the prior
range of conditioning parameters, finding that the recovery varies little across the prior
(\S\ref{sec:results:cond}). The reconstruction fidelity drops only in the highly ionized windows.
We find that the fidelity of the reconstruction, including this drop, is set by the autoencoder,
which smooths the small neutral islands left near the end of reionization, and that the denoiser
delivers $\sim 100\%$ of what the frozen autoencoder retains (\S\ref{sec:results:ceiling}).
Since autoencoder training involved only the HR field, a coarser LR input needs only a
finetuned denoiser (\S\ref{sec:method:training}). At $16\times$, the finetuned denoiser reaches the
same reconstruction fidelity as the $8\times$ model (\S\ref{sec:results:x16}). However, on a new
realization of the matter field, the frozen model degrades even below the input Nyquist
frequency. Finetuning both stages on $600$ lightcones of the new realization, $10\%$ of the original training set, restores near-ceiling
coherence and power, but at the expense of fidelity on the original realization
(\S\ref{sec:results:newseed}).

The reconstruction is inexpensive compared to the HR simulation. Generating one HR lightcone with
\texttt{SCRIPT} requires approximately $2$ hours on a single CPU core, while the \SuperSCRIPT{}
reconstruction takes $4$ seconds on one A100 GPU and $23$ minutes on a single CPU core. The
end-to-end speed-up on the GPU, including the LR run, is roughly $560\times$. An LR lightcone also
has $512$ times fewer voxels than HR. LR ensembles can then be stored across the prior, and selected HR volumes reconstructed on demand as $21$-cm, kSZ and optical-depth maps for non-Gaussian
statistics and map-level auto- and cross-correlations. Repeated forward evaluations also become
feasible on a given matter realization, for source-model exploration, cross-correlations with
tracers drawn from the same simulation, and parameter derivatives for forecasts and inference.
These evaluations fit directly into existing analyses, since the model uses the same source
framework as earlier studies of CMB constraints on reionization~\citep{Jain2023framework,
Jain2024kSZ, Jain2024tau} and the same \texttt{SCRIPT} forward model as parameter-inference studies
for upcoming $21$-cm measurements~\citep{Maity2023, ChoudhuryParanjapeMaity2024}.

Since the autoencoder learns the morphology distribution without source-parameter labels,
which enter only through the diffusion conditioning, it can be reused across source
parameterizations within that distribution. Sampling cosmic variance, however, currently requires a finetuned model for each matter realization, since the model inputs lack the unresolved density field that sets the ionization structure (\S\ref{sec:results:newseed}). Per-realization finetuning costs $10\%$ of the original training set in HR simulations. Conditioning on the density field, together with training on several realizations, should alleviate
this requirement. This input adds no simulation cost, since the
$21$-cm and kSZ maps of each realization require the density field. In future work, we
will investigate this conditioning along with a latent transport that accepts a variable
super-resolution factor. This approach will likely enable a single trained model to generalize
across different matter realizations and LR resolutions.
\acknowledgments

DJ thanks Barun Maity for useful discussions. DJ and CH acknowledge funding from the Volkswagen
Foundation, and support through the Federal Ministry of Education and Research (BMBF) and the Ministry of Science, Research and the Arts of Baden-Württemberg. This work was supported by the Deutsche Forschungsgemeinschaft (DFG, German Research
Foundation) under Germany's Excellence Strategy EXC~2181/1 -- 390900948 (the Heidelberg STRUCTURES
Excellence Cluster). TRC acknowledges support from the Department of Atomic Energy, Government of
India, under project no.~12-R\&D-TFR-5.02-0700. CH also acknowledges funding via the SciFM consortium (05D25VH3) funded by the German Federal Ministry of Research, Technology, and Space (BMFTR) in the ErUM-Data action plan. The authors acknowledge support by the High
Performance and Cloud Computing Group at the Zentrum f\"ur Datenverarbeitung of the University of
T\"ubingen, the state of Baden-W\"urttemberg through bwHPC and the German Research Foundation (DFG)
for funding under Project number 455787709 (bwForCluster BinAC~2). The \texttt{SuperSCRIPT} code
builds on the \texttt{map2map}\footnote{\url{https://github.com/eelregit/map2map}}
framework~\citep{Li2021}. DJ used generative AI tools (Kimi K3, Moonshot AI; GLM-5.3, Z.ai; Claude,
Anthropic) to assist with code refactoring and review, and text editing. All AI-assisted
contributions were checked and verified by DJ.

\section*{Code Availability Statement}

The \texttt{SuperSCRIPT} code is available from the corresponding author upon reasonable request and will be made publicly available in a future release.

\section*{Data Availability Statement}

The simulated LR lightcones, together with the trained autoencoder and denoiser weights, are available from the corresponding author upon reasonable request.

\appendix
\counterwithin{figure}{section}
\counterwithin{table}{section}
\section{Reconstruction fidelity of the selected lightcones}
\label{app:lightcones}

\begin{figure}[!ht]
\centering
\includegraphics[width=\columnwidth]{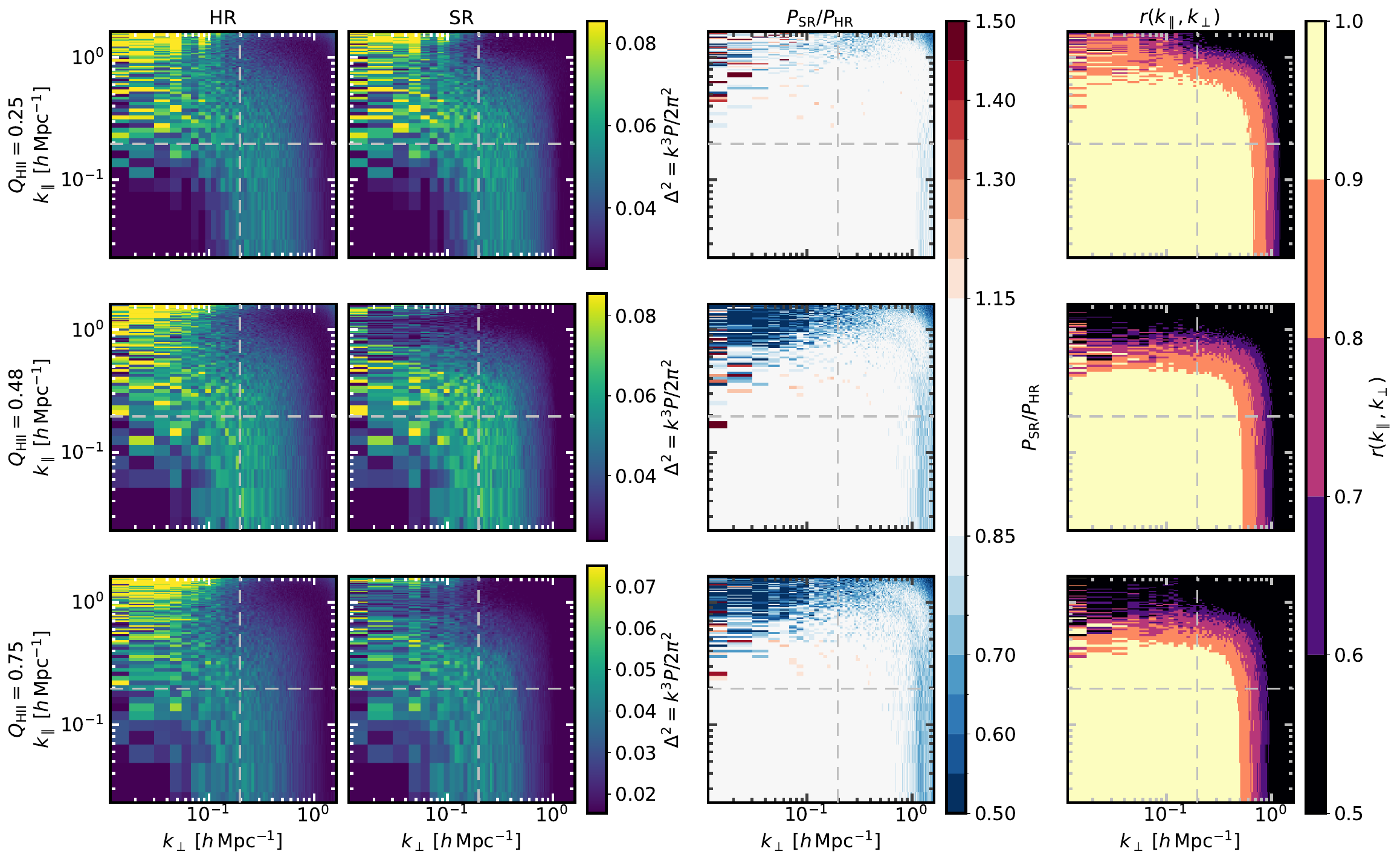}
\caption{Cylindrical spectra of the lightcone with optical depth $\tau=0.040$ at three stages of
reionization. \emph{Columns:} the dimensionless cylindrical power $\Delta^2$ of the HR and SR
fields, their power ratio and their cross-coherence. White cells mark $0.85\le P_{\rm SR}/P_{\rm
HR}\le1.15$ and dashed lines the LR Nyquist frequency.}
\label{fig:cyl:s1}
\end{figure}

\begin{figure}
\centering
\includegraphics[width=\columnwidth]{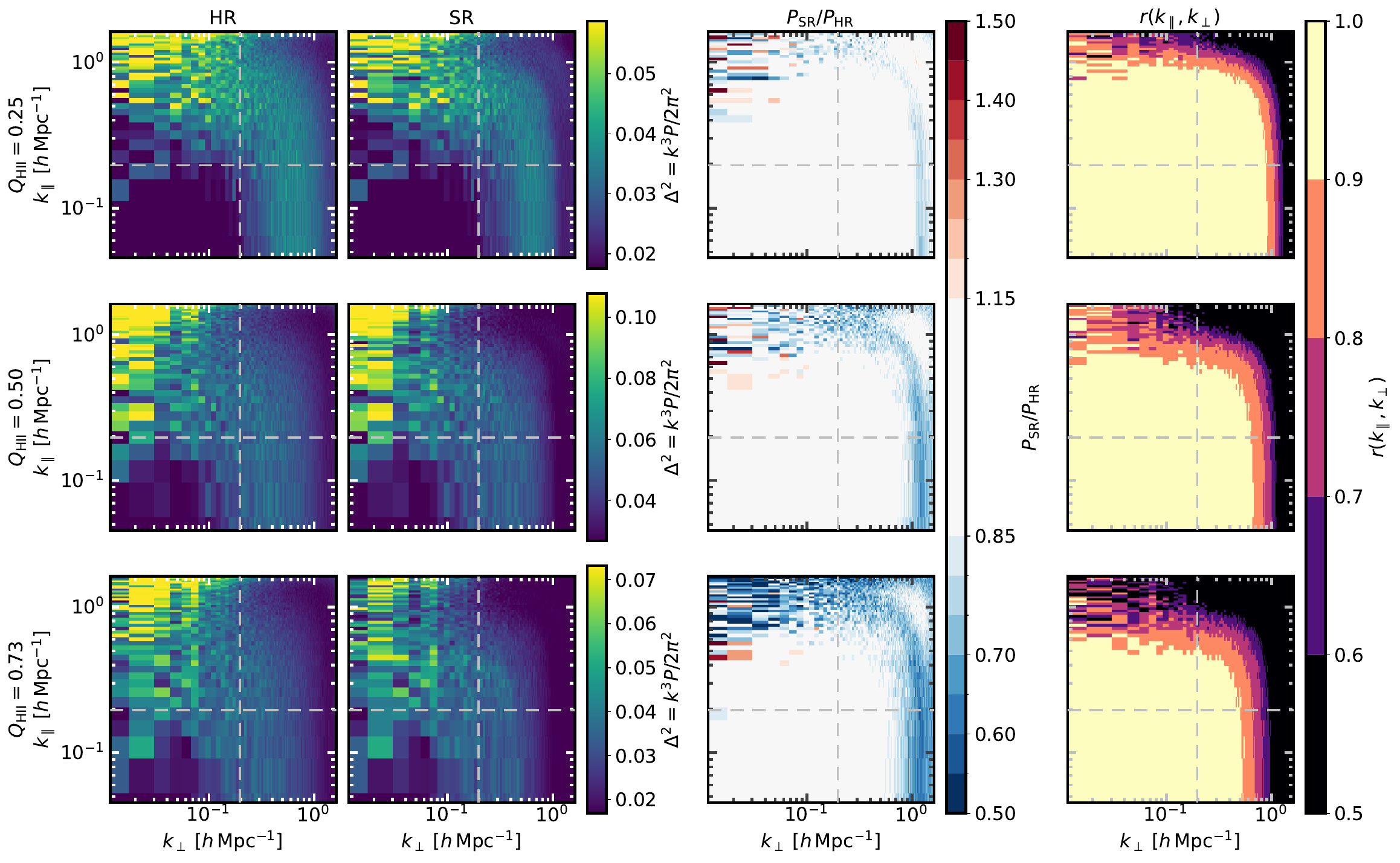}
\caption{Cylindrical spectra of the lightcone with optical depth $\tau=0.081$ at three stages of
reionization. \emph{Columns:} the dimensionless cylindrical power $\Delta^2$ of the HR and SR
fields, their power ratio and their cross-coherence. White cells mark $0.85\le P_{\rm SR}/P_{\rm
HR}\le1.15$ and dashed lines the LR Nyquist frequency.}
\label{fig:cyl:s3}
\end{figure}

\begin{figure}
\centering
\includegraphics[width=\columnwidth]{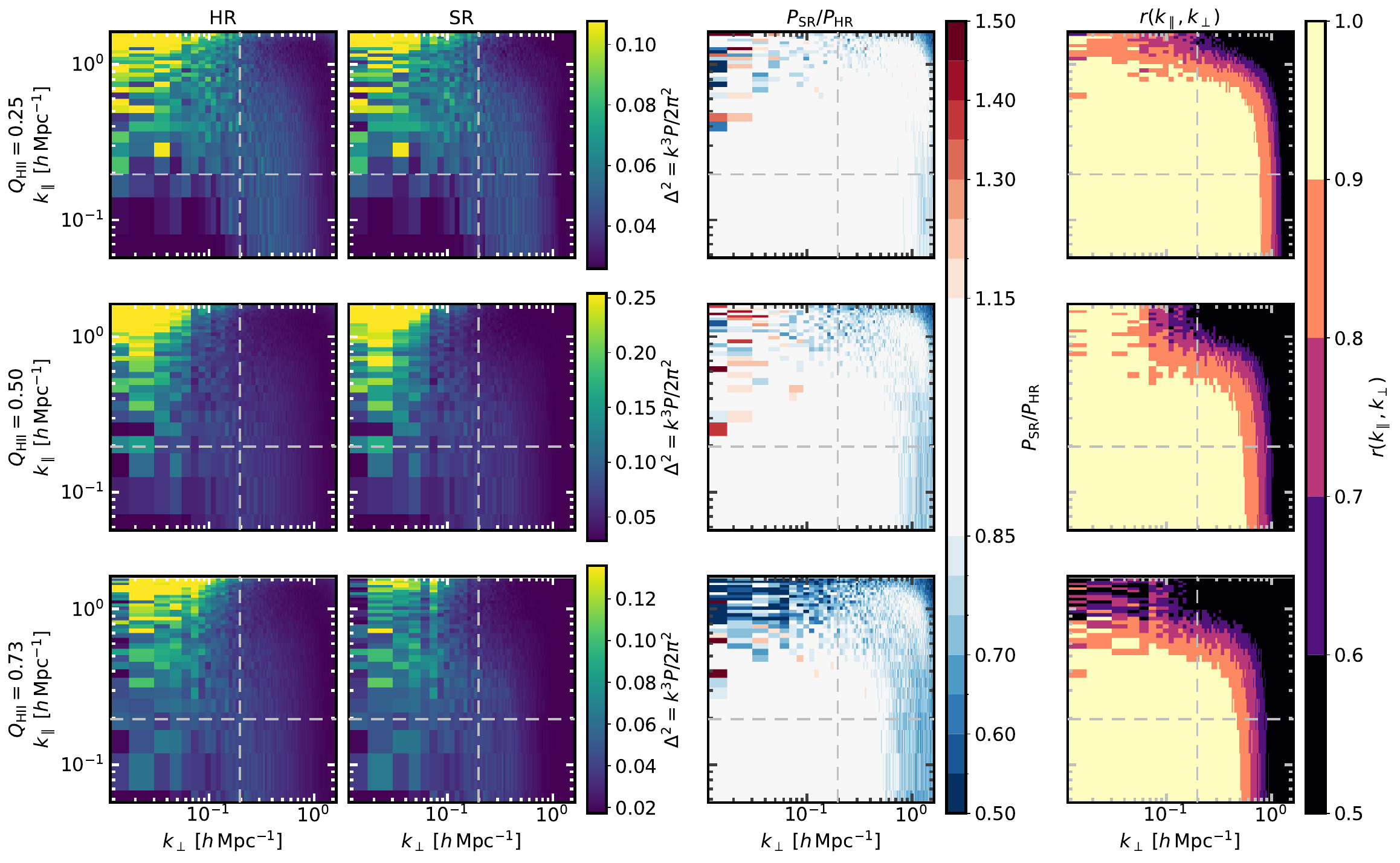}
\caption{Cylindrical spectra of the lightcone with optical depth $\tau=0.102$ at three stages of
reionization. \emph{Columns:} the dimensionless cylindrical power $\Delta^2$ of the HR and SR
fields, their power ratio and their cross-coherence. White cells mark $0.85\le P_{\rm SR}/P_{\rm
HR}\le1.15$ and dashed lines the LR Nyquist frequency.}
\label{fig:cyl:s4}
\end{figure}

\begin{figure}
\centering
\includegraphics[width=0.9\columnwidth]{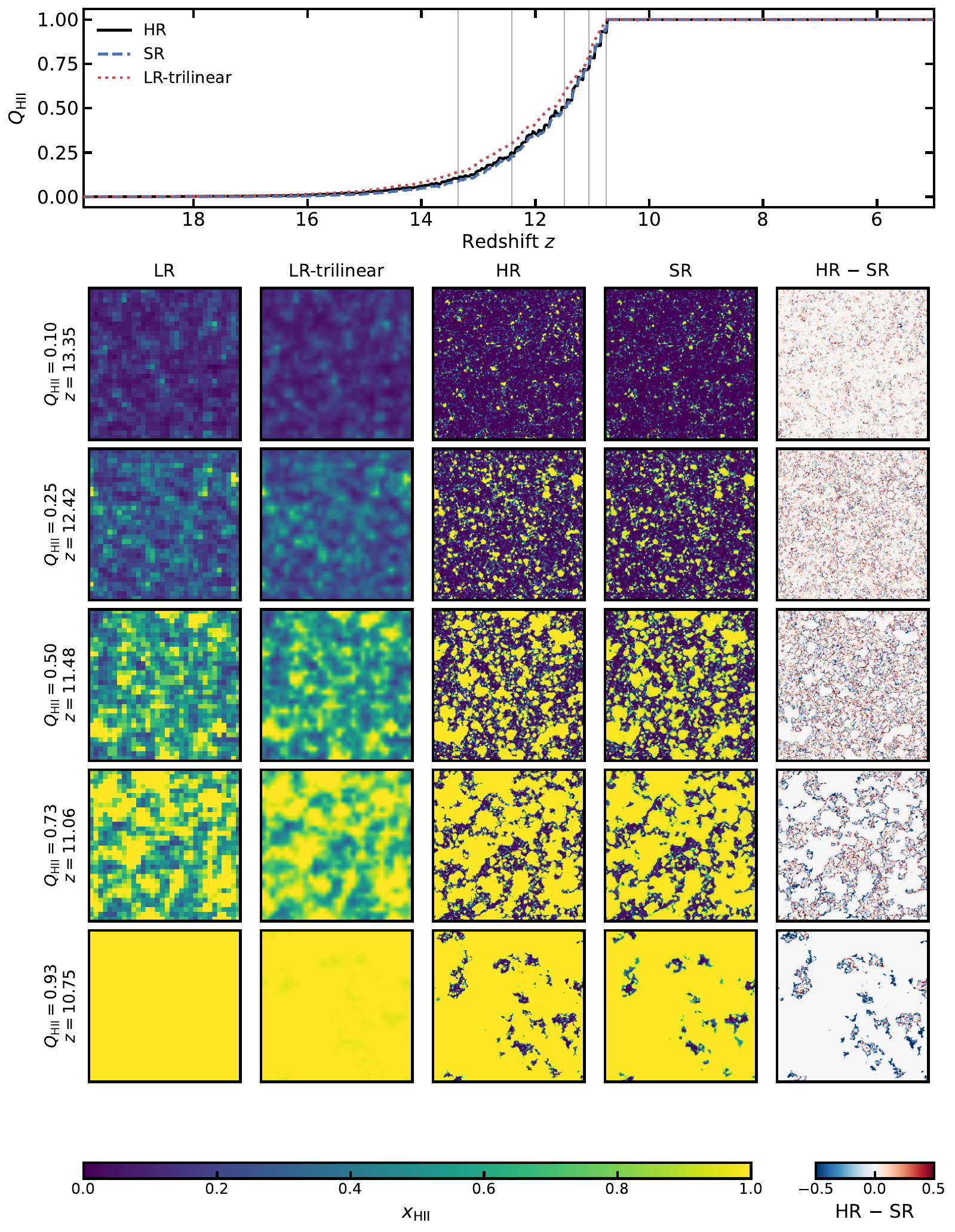}
\caption{Reconstruction of the lightcone with optical depth $\tau=0.102$. The top panel shows the
ionization histories $Q_{\rm HII}(z)$ of HR, SR and LR-trilinear, with vertical lines marking the
redshifts of the slices. The bottom panel shows transverse slices of LR, LR-trilinear, HR and SR at
five stages of reionization, followed by the residual $x_{\rm HII}^{\rm HR-SR}$.}
\label{fig:lc:high}
\end{figure}

\begin{figure}
\centering
\includegraphics[width=\columnwidth]{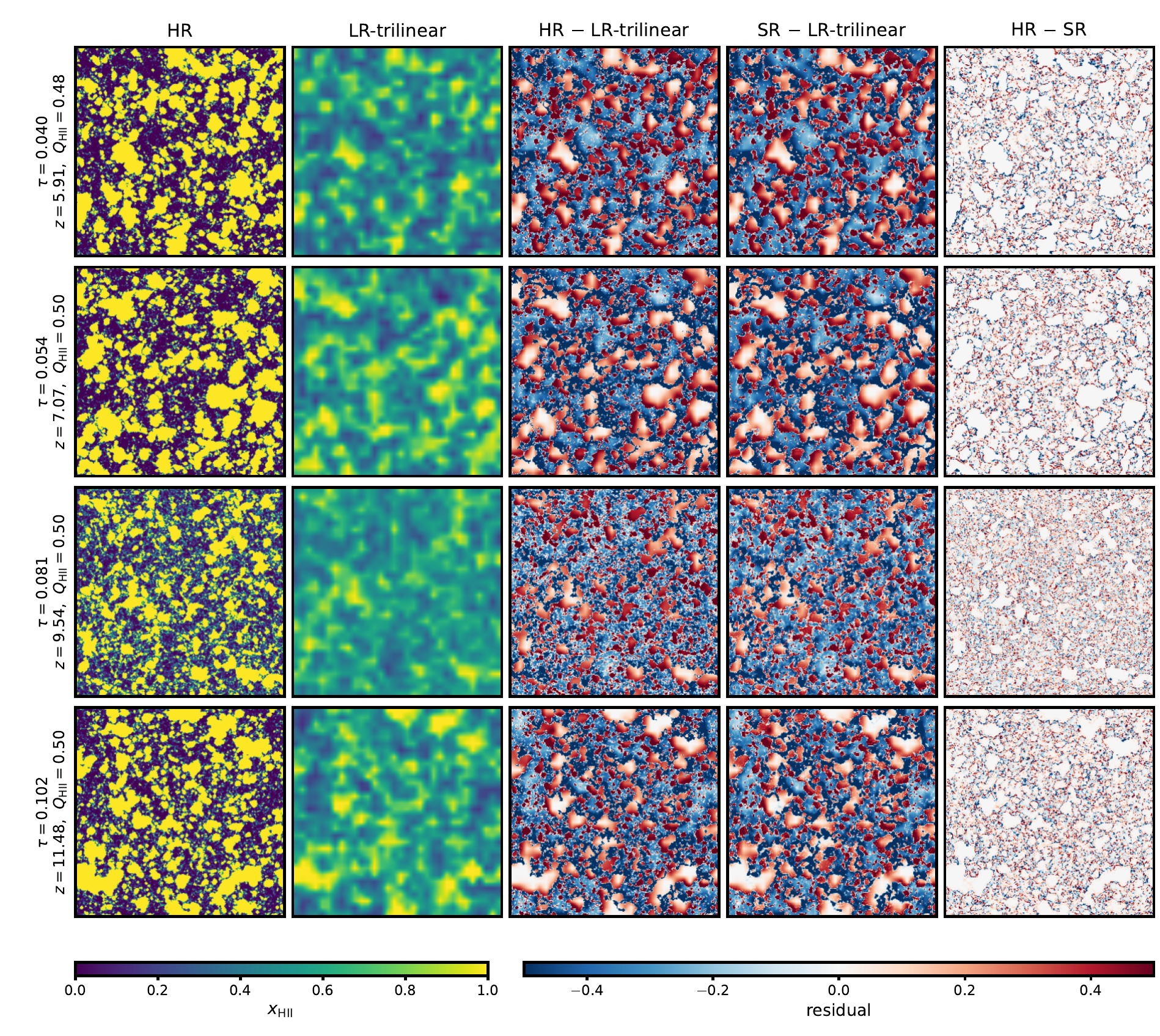}
\caption{$x_{\rm HII}$ slices and $x_{\rm HII}$ residual slices for the four lightcones with optical
depths $\tau=0.040$, $0.054$, $0.081$ and $0.102$ at their midpoint of reionization.
\emph{Columns:} HR; LR-trilinear; $x_{\rm HII}^{\rm HR-tri}$; $x_{\rm HII}^{\rm SR-tri}$; $x_{\rm
HII}^{\rm HR-SR}$, where tri denotes LR-trilinear.}
\label{fig:residual}
\end{figure}

This appendix presents the figures of the four lightcones of \S\ref{sec:results:qual} as a
reference for the recovery across reionization histories, from the late and rapid reionization of
the $\tau=0.040$ lightcone to the early reionization of the $\tau=0.102$ lightcone, and
Table~\ref{tab:cyl:showcase} lists their midpoints and durations.
Figures~\ref{fig:cyl:s1}--\ref{fig:cyl:s4} show the cylindrical spectra of the other three in the
layout of Figure~\ref{fig:cyl:s2}. To show an extreme reionization history, we also present in Figure~\ref{fig:lc:high} the transverse slices of the
$\tau=0.102$ lightcone in the layout of Figure~\ref{fig:lc:planck}. Figure~\ref{fig:residual} shows the
residuals of HR and SR with respect to LR-trilinear for all four at their midpoint of reionization,
which isolate the structure that the model must generate. The HR and SR residuals share the same
bubbles at every scale, and their difference is confined to voxel-wide features along the bubble
boundaries.

\section{Recovery of density-weighted fields}
\label{app:density}

\begin{figure}[!ht]
\centering
\includegraphics[width=0.9\columnwidth]{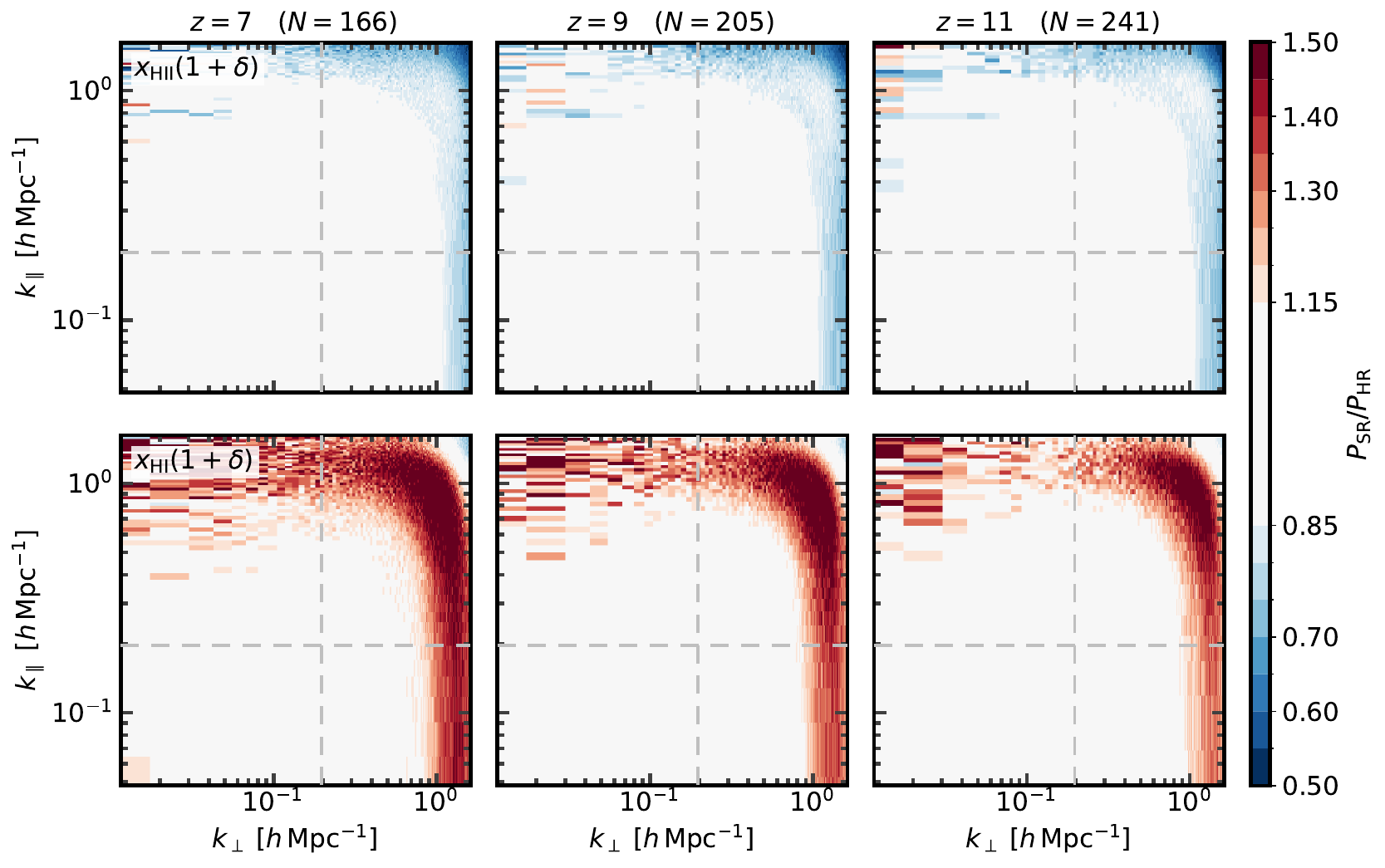}
\includegraphics[width=0.9\columnwidth]{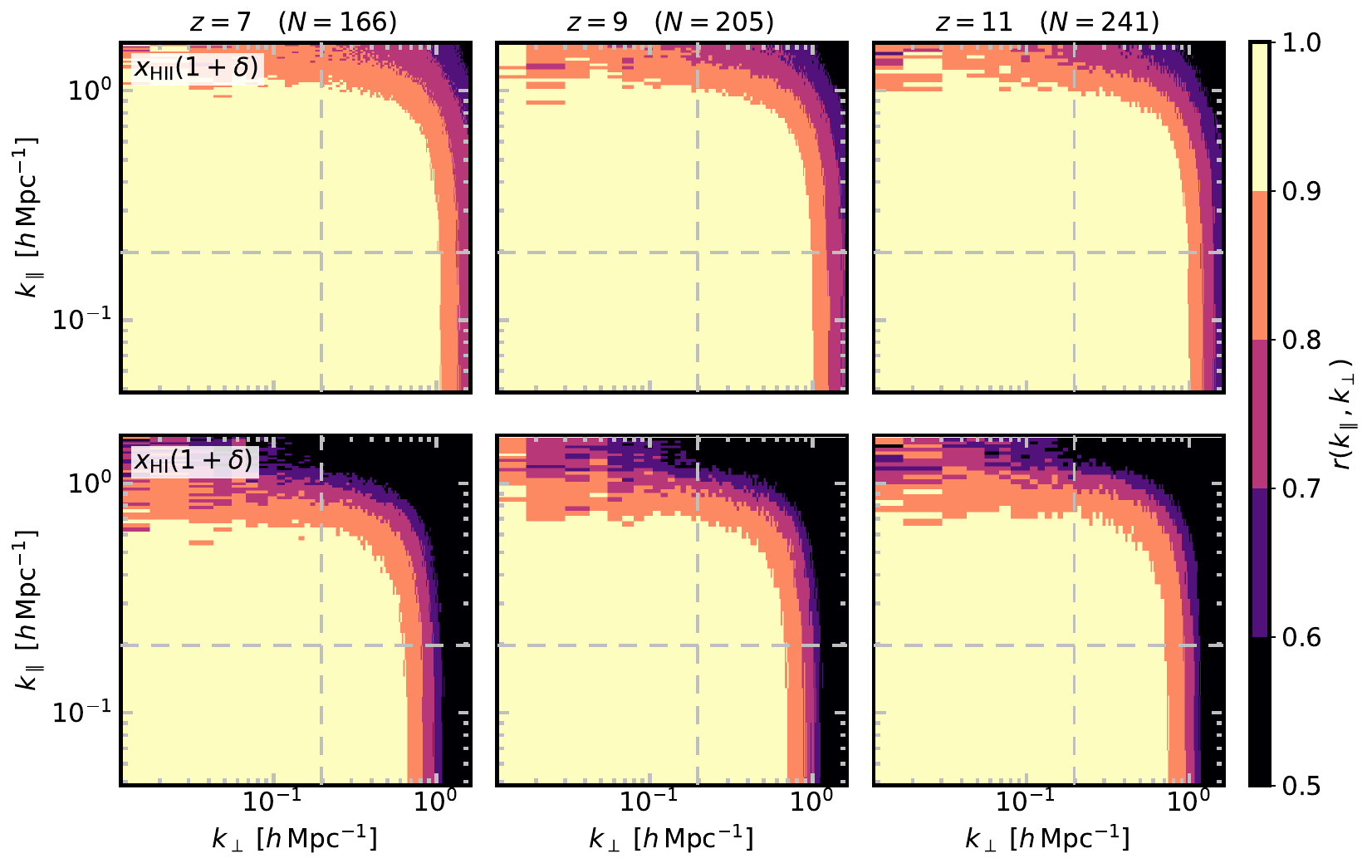}
\caption{Population-median cylindrical power ratio $P_{\rm SR}/P_{\rm HR}$ (top two rows) and
cross-coherence $r(k_\parallel,k_\perp)$ (bottom two rows) of the density-weighted fields in the
$(k_\parallel,k_\perp)$ plane at $z=7$, $9$ and $11$, for the ionized field $x_{\rm HII}(1+\delta)$
and then the neutral field $x_{\rm HI}(1+\delta)$, over the $N$ lightcones partially ionized in each
$w_z=1$ window. HR and SR are multiplied by the same matter density. White cells mark $0.85\le
P_{\rm SR}/P_{\rm HR}\le1.15$ and dashed lines the LR Nyquist frequency.}
\label{fig:cyl:popdens}
\end{figure}

\begin{figure}
\centering
\includegraphics[width=\columnwidth]{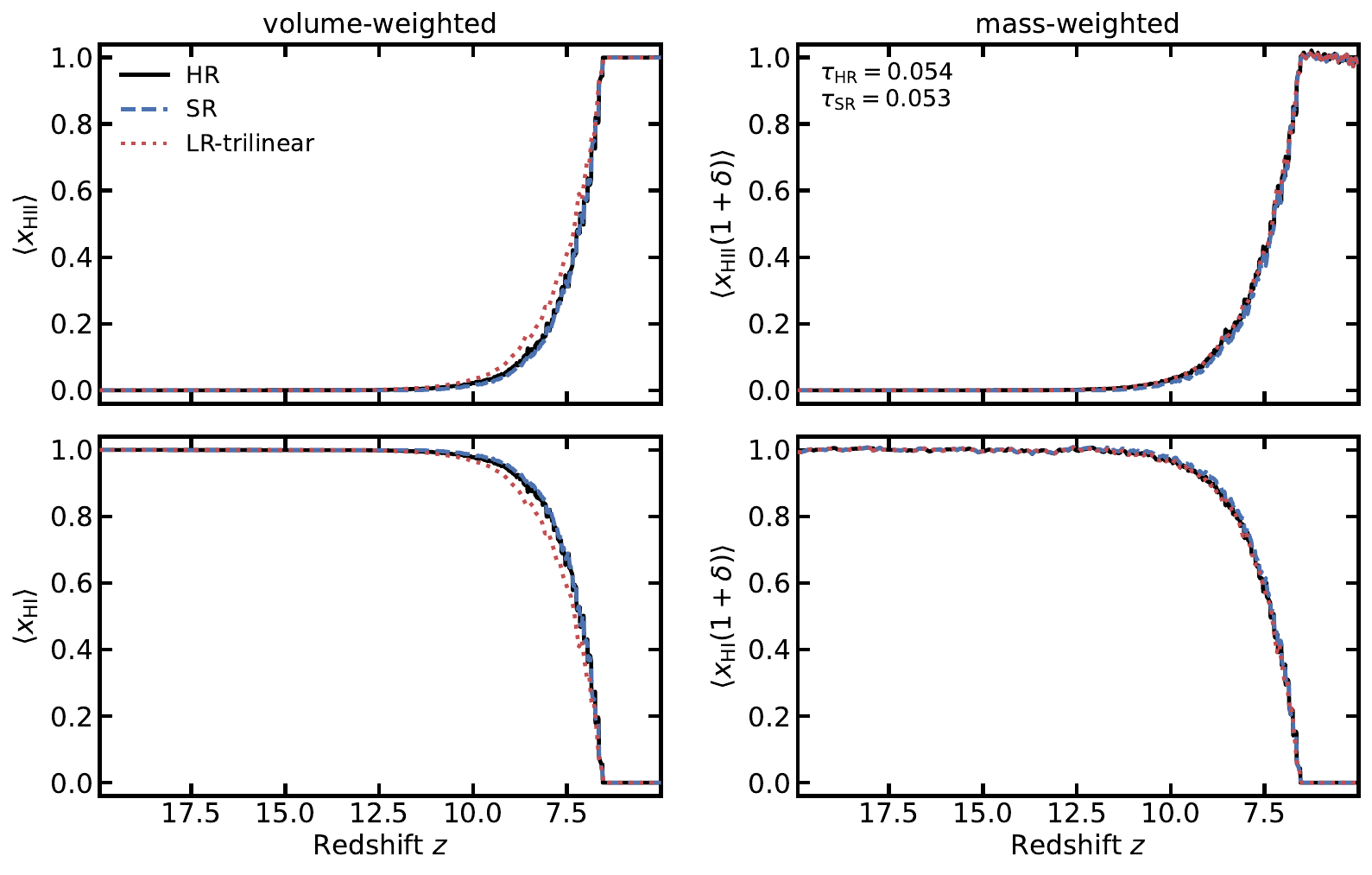}
\caption{Ionization histories of the lightcone with optical depth $\tau=0.054$ for HR, SR and
LR-trilinear. \emph{Top:} the ionized fraction, volume-weighted $\langle x_{\rm HII}\rangle$ and
mass-weighted $\langle x_{\rm HII}(1+\delta)\rangle$. \emph{Bottom:} the neutral fractions
$\langle x_{\rm HI}\rangle$ and $\langle x_{\rm HI}(1+\delta)\rangle$.}
\label{fig:qmass}
\end{figure}

Here we present the recovery of the density-weighted fields $x_{\rm HII}(1+\delta)$ and
$x_{\rm HI}(1+\delta)=(1-x_{\rm HII})(1+\delta)$, on the same $600$ test lightcones and the same
windows as Section~\ref{sec:results}. Here $\delta$ is the 2LPT dark-matter density contrast of
the common matter realization on the HR grid (\S\ref{sec:data:sims}). Figure~\ref{fig:cyl:popdens}
shows the population-median power ratio and coherence of the two fields. Within half the HR Nyquist
frequency SR recovers $0.96$--$0.99$ of the HR power of $x_{\rm HII}(1+\delta)$ at a coherence of
$0.98$, and $1.05$--$1.13$ of the power of $x_{\rm HI}(1+\delta)$ at a coherence of
$0.91$--$0.93$. We also present in Figure~\ref{fig:qmass} the volume- and mass-weighted ionized
and neutral fractions of the $\tau=0.054$ lightcone.

\section{Selected lightcones at \texorpdfstring{$16\times$}{16x}}
\label{app:x16}

\begin{figure}[!ht]
\centering
\includegraphics[width=\columnwidth]{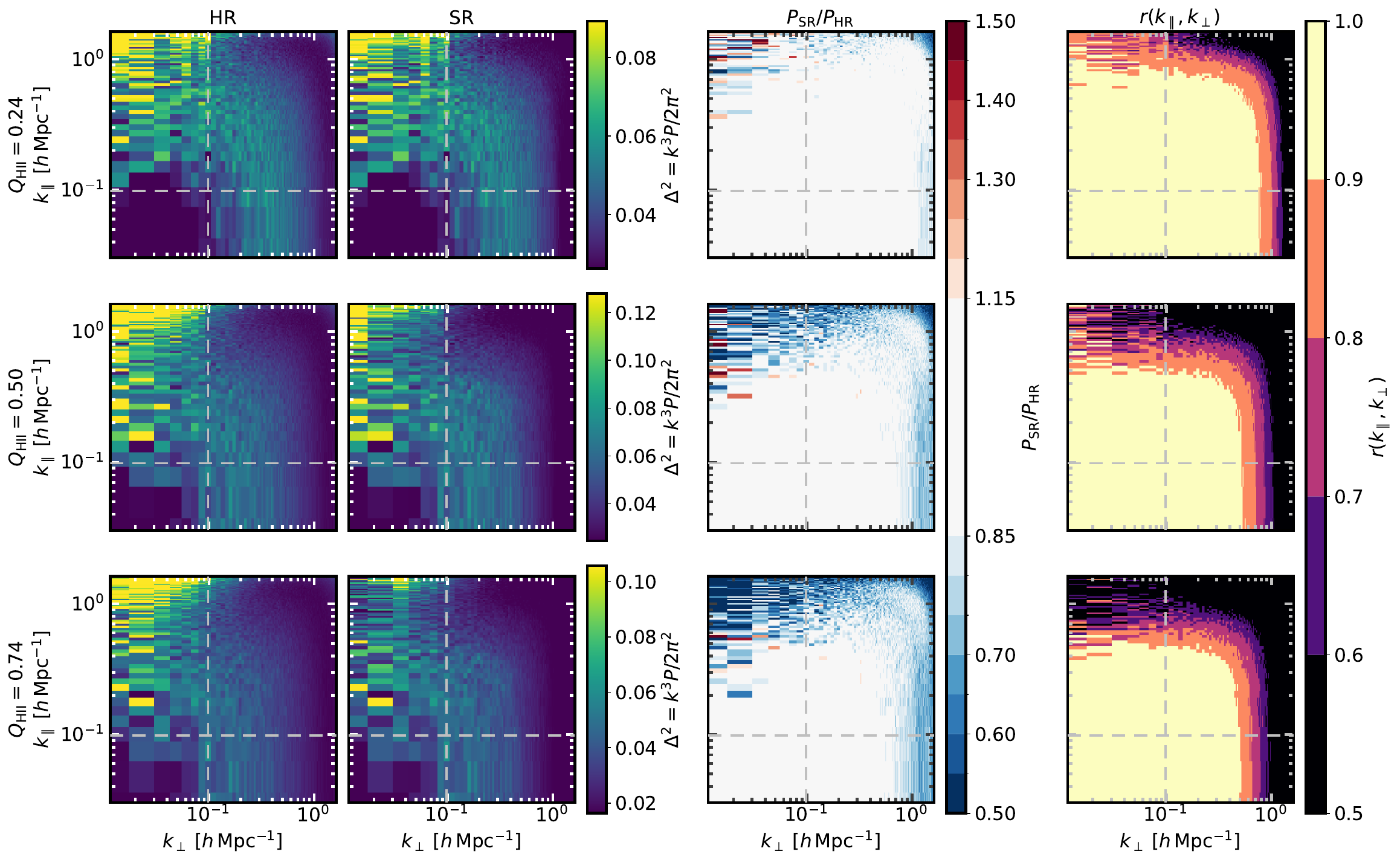}
\caption{Cylindrical spectra of the lightcone with optical depth $\tau=0.054$ at three stages of
reionization, for the finetuned $16\times$ model. \emph{Columns:} the dimensionless cylindrical
power $\Delta^2$ of the HR and SR fields, their power ratio and their cross-coherence. White cells
mark $0.85\le P_{\rm SR}/P_{\rm HR}\le1.15$ and dashed lines the LR Nyquist frequency at
$16\times$.}
\label{fig:cyl:s3g16rt}
\end{figure}

\begin{figure}
\centering
\includegraphics[width=\columnwidth]{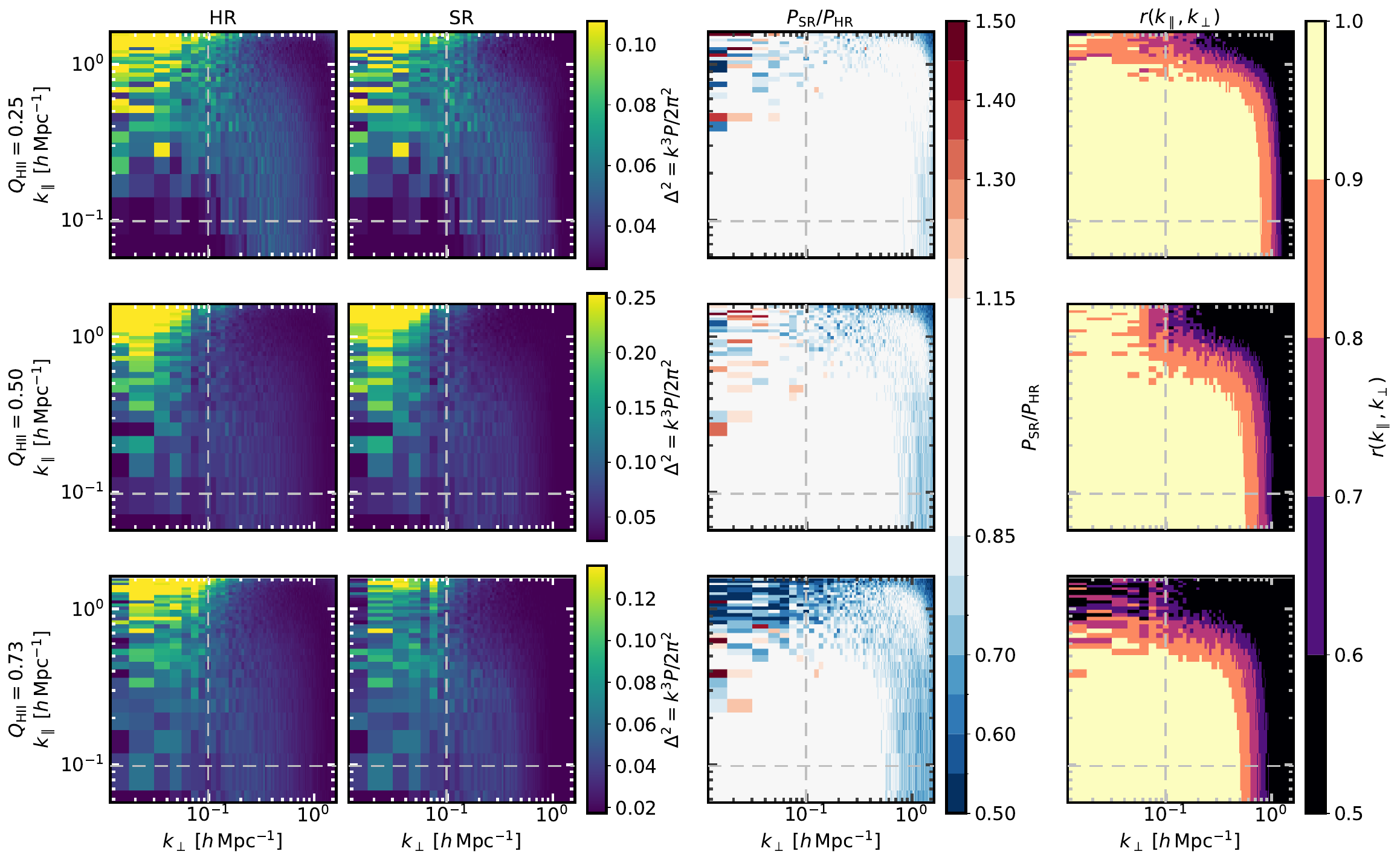}
\caption{Cylindrical spectra of the lightcone with optical depth $\tau=0.102$ at three stages of
reionization, for the finetuned $16\times$ model. \emph{Columns:} the dimensionless cylindrical
power $\Delta^2$ of the HR and SR fields, their power ratio and their cross-coherence. White cells
mark $0.85\le P_{\rm SR}/P_{\rm HR}\le1.15$ and dashed lines the LR Nyquist frequency at
$16\times$.}
\label{fig:cyl:s2g16}
\end{figure}

We present the finetuned $16\times$ model for two of the selected lightcones, the Planck-like
$\tau=0.054$ lightcone and the $\tau=0.102$ lightcone, which reionizes far earlier than the Planck
measurement allows. Figures~\ref{fig:cyl:s3g16rt} and~\ref{fig:cyl:s2g16} show their cylindrical
spectra, consistent with the $8\times$ model as in Table~\ref{tab:x16pop}.

\section{Selected lightcones on a new matter realization}
\label{app:newseed}

\begin{figure}[!ht]
\centering
\includegraphics[width=\columnwidth]{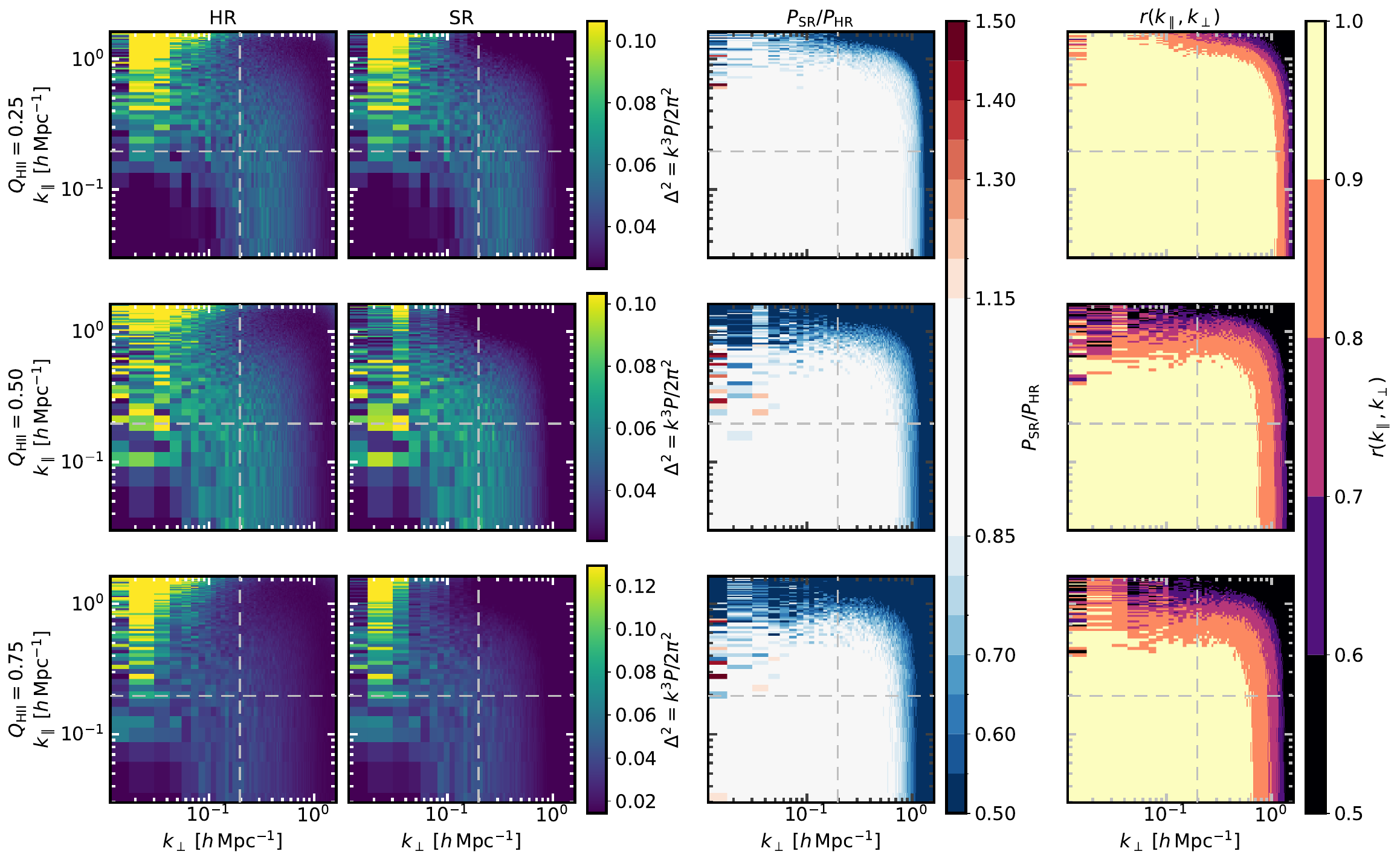}
\caption{Cylindrical spectra of the lightcone with optical depth $\tau=0.054$ on the new
matter realization, after the finetune, at three stages of reionization. \emph{Columns:} the
dimensionless cylindrical power $\Delta^2$ of the HR and SR fields, their power ratio and their
cross-coherence. White cells mark $0.85\le P_{\rm SR}/P_{\rm HR}\le1.15$ and dashed lines the LR
Nyquist frequency.}
\label{fig:newseed:s3}
\end{figure}

\begin{figure}
\centering
\includegraphics[width=\columnwidth]{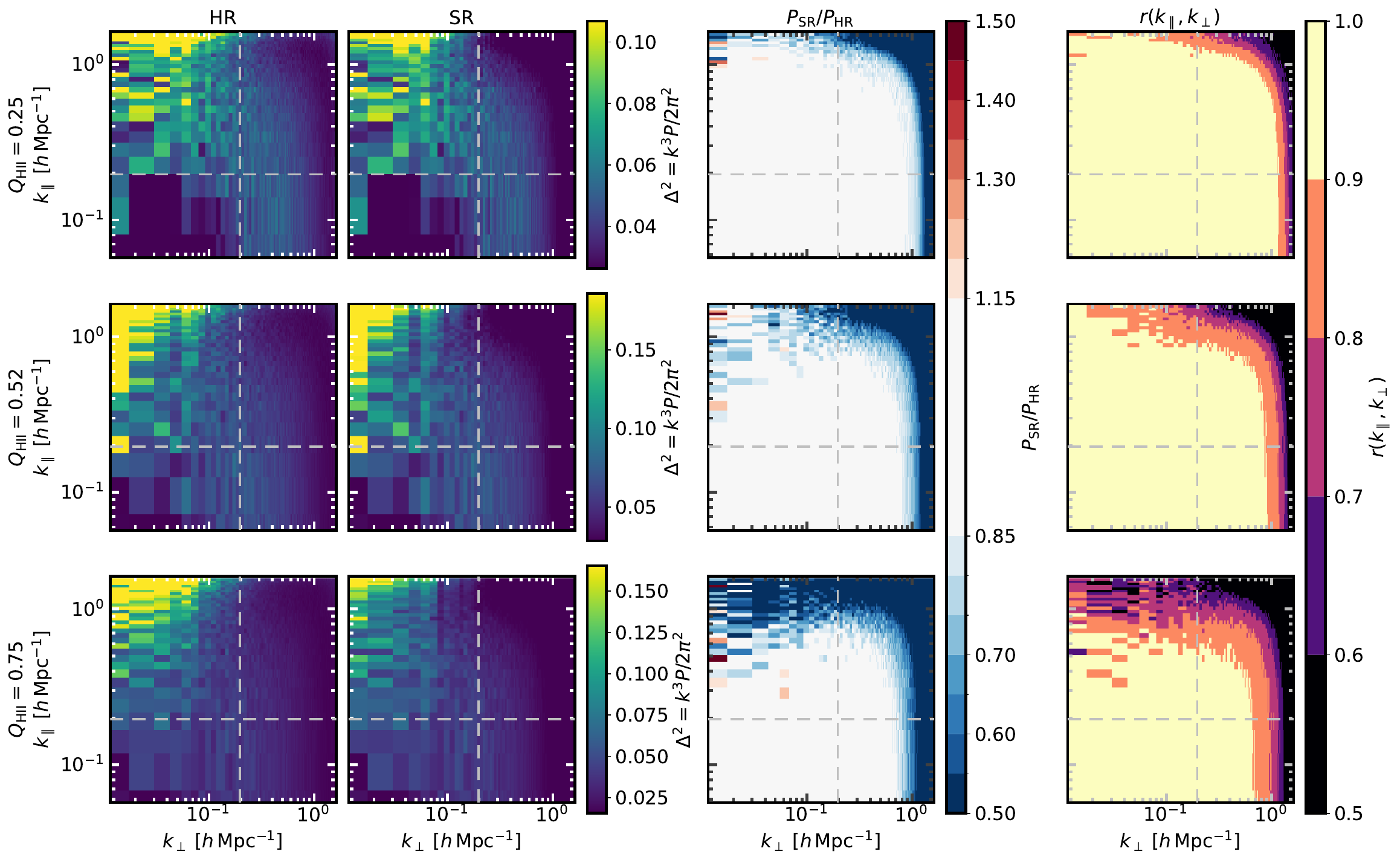}
\caption{Cylindrical spectra of the lightcone with optical depth $\tau=0.102$ on the new
matter realization, after the finetune, at three stages of reionization. \emph{Columns:} the
dimensionless cylindrical power $\Delta^2$ of the HR and SR fields, their power ratio and their
cross-coherence. White cells mark $0.85\le P_{\rm SR}/P_{\rm HR}\le1.15$ and dashed lines the LR
Nyquist frequency.}
\label{fig:newseed:s2}
\end{figure}

We show the finetuned model on the new matter realization for two of the selected lightcones,
the Planck-like $\tau=0.054$ lightcone and the $\tau=0.102$ lightcone, which reionizes far earlier
than the Planck measurement allows. Figures~\ref{fig:newseed:s3} and~\ref{fig:newseed:s2} show their
cylindrical spectra after the finetune of \S\ref{sec:results:newseed}.

\clearpage

\bibliographystyle{JHEP}
\bibliography{references}

\end{document}